%% file: main.tex
\documentclass[letterpaper]{article} 
\usepackage{aaai2026}  
\usepackage{times}  
\usepackage{helvet}  
\usepackage{courier}  
\usepackage[hyphens]{url}  
\usepackage{graphicx} 
\usepackage{natbib}  
\usepackage{caption} 
\usepackage{algorithm}
\usepackage{xcolor}
\usepackage{newfloat}
\usepackage{listings}
\DeclareCaptionStyle{ruled}{labelfont=normalfont,labelsep=colon,strut=off} 
\floatstyle{ruled}
\newfloat{listing}{tb}{lst}{}
\floatname{listing}{Listing}
\usepackage{enumitem}
\usepackage{amsmath, amssymb, amsthm}
\usepackage{amsfonts}
\usepackage{newtxmath}
\usepackage{lineno}
\usepackage{algpseudocode}
\usepackage{graphicx}
\usepackage{multirow}
\usepackage{subcaption} 

\newtheorem{theorem}{Theorem}
\newtheorem{lemma}{Lemma}
\newtheorem{assumption}{Assumption}

\title{Federated Vision-Language-Recommendation with Personalized Fusion}

\author{
    Zhiwei Li\textsuperscript{\rm 1},
    Guodong Long\textsuperscript{\rm 1},
    Jing Jiang\textsuperscript{\rm 1},
    Chengqi Zhang\textsuperscript{\rm 2},
    Qiang Yang\textsuperscript{\rm 2}
}
\affiliations{
    \textsuperscript{\rm 1} Australian AI Institute, Faculty of Engineering and IT, University of Technology Sydney \\
    \textsuperscript{\rm 2} Department of Data Science and Artificial Intelligence, The Hong Kong Polytechnic University\\
    zhw.li@outlook.com,
    \{guodong.long, jing.jiang\}@uts.edu.au,
    \{chengqi.zhang, profqiang.yang\}@polyu.edu.hk
}

\usepackage{bibentry}

\begin{document}

\maketitle

\begin{abstract}
Applying large pre-trained Vision-Language Models to recommendation is a burgeoning field, a direction we term Vision-Language-Recommendation (VLR). 
Bringing VLR to user-oriented on-device intelligence within a federated learning framework is a crucial step for enhancing user privacy and delivering personalized experiences.
This paper introduces FedVLR, a federated VLR framework specially designed for user-specific personalized fusion of vision-language representations.
At its core is a novel bi-level fusion mechanism:
The server-side multi-view fusion module first generates a diverse set of pre-fused multimodal views. 
Subsequently, each client employs a user-specific mixture-of-expert mechanism to adaptively integrate these views based on individual user interaction history.
This designed lightweight personalized fusion module provides an efficient solution to implement a federated VLR system. 
The effectiveness of our proposed FedVLR has been validated on seven benchmark datasets.
\end{abstract}

%
\begin{links}
    \link{Code}{https://github.com/mtics/FedVLR}
\end{links}

\section{Introduction}

Vision-Language Models (VLMs) are pushing the boundaries of personalized recommendation by interpreting the rich content of items~\cite{wei2024towards}, a direction we conceptualize as Vision-Language-Recommendation (VLR). 
By understanding visual aesthetics and textual semantics, VLR models can move beyond simple item identifiers (IDs) to capture a deeper, more nuanced understanding of user preferences~\cite{zhou2023comprehensive}. 
Deploying these powerful VLR models directly on user devices is a significant step forward, as this on-device approach enhances user privacy, reduces network latency, and grants users direct ownership of their data~\cite{yin2024device}, aligning with current privacy-centric principles~\cite{voigt2017eu}.

This imperative has catalyzed the burgeoning field of Federated Vision-Language Models (FedVLMs), which aims to train powerful VLMs on decentralized data without compromising privacy~\cite{ren2024advances}. 
Initial research in FedVLMs has primarily focused on foundational challenges, such as adapting large model architectures to the federated setting~\cite{liu2020federated}, and mitigating prohibitive communication costs~\cite{guo2023promptfl}.
While these efforts are vital, they often overlook a more subtle challenge specific to the on-device recommendation tasks: how to fuse vision-language signals in an user-oriented personalized way.


Consider the process of choosing a movie. What factors contribute to the decision? 
One user might be attracted by a beautiful movie poster, another might be impacted by the story described in the text summary, while a third might rely on collaborative signals from friends with similar tastes~\cite{liu2024mmrec,liu2024multimodal}. 
These factors, spanning visual, textual, and collaborative signals, contribute to each user's final decision in a highly individualized manner~\cite{wei2024towards}, and points to a challenge deeper than just the statistical heterogeneity of data, i.e., \emph{preference heterogeneity}. 
We define this as the phenomenon where users exhibit diverse and personal criteria when evaluating and weighing information from different modalities. 
Inspired by this, we believe that a personalized multi-modal fusion module is the critical component needed to enhance federated VLR by capturing these fine-grained user preferences across all modalities.

This diversity means that a single and one-size-fits-all module for fusing multimodal signals is inherently suboptimal. 
Yet, existing federated recommendation systems often fail to address this. 
They are either content-agnostic, relying only on interaction IDs~\cite{lin2020fedrec,zhang2023dual}, or they impose a globally uniform fusion logic on all users~\cite{li2024towards,feng2024robust}. 
They neglect the critical need for the fusion module itself to be personalized.

\begin{figure}
    \centering
    \includegraphics[width=0.95\linewidth]{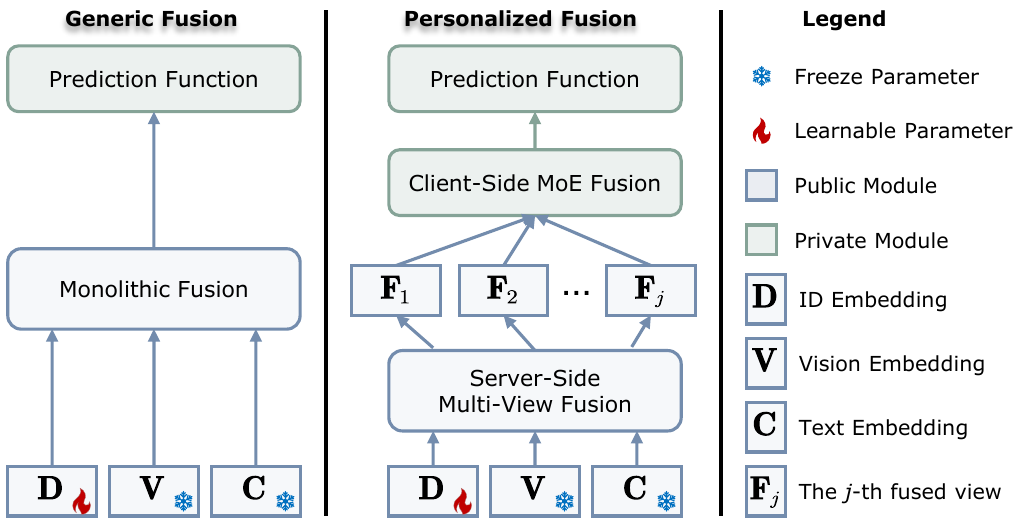}
    \caption{
    Paradigm shift from a monolithic generic fusion (Left) to our personalized fusion (Right), enabling fine-grained on-device personalization by decoupling server-side view generation from client-side refinement.
    }
    \label{fig:compare}
\end{figure}

To address this gap, we propose a novel \textbf{Fed}erated \textbf{V}ision-\textbf{L}anguage-\textbf{R}ecommendation framework with Personalized Fusion (\textbf{FedVLR}). 
Our framework learns under a standard federated recommendation setting, where all item features are stored on the server, while user interaction histories remain privately on each client's device~\cite{zhang2023dual}, respecting data ownership and mitigates privacy risks. 
The core innovation lies in a personalized and dynamic fusion strategy, realized through our proposed Bi-Level Fusion Mechanism (BLFM). 
As shown in Fig.~\ref{fig:compare}, instead of imposing a generic fusion logic on all users through a shared single module (Left), our BLFM enables personalized fusion by decoupling the fusion into two levels (Right). 
The server firstly generates multiple feature views using diverse operators, handling the major computational load.
Subsequently, each client learns a lightweight, personalized refinement over these views using a Mixture-of-Experts (MoE) module based on user's private history.
This architecture maximizes the utility of server-side operators to deliver personalization in a lightweight manner, effectively addressing preference heterogeneity in the federated VLR settings.

Our main contributions are summarized as follows:
\begin{itemize}

\item We are the first to formalize and tackle the personalized multimodal fusion problem in federated VLR settings.

\item Our proposed Bi-Level Fusion Mechanism contributes a new paradigm for modality fusion that can be generalized to the broader domain of multimodal learning, enabling fine‐grained multimodal personalization.

\item We design FedVLR as a versatile framework that seamlessly enhances a wide range of existing ID-based federated models with content-aware personalization.

\item Extensive empirical validation shows that FedVLR not only substantially improves existing baselines but can also outperform centralized models in certain low-data regimes. Source code is provided for reproducibility.











\end{itemize}

\section{Related Work}
\label{sec:related_work}

\subsection{Vision-Language-Recommendation Models}

A prominent direction in representation learning is translating multimodal perception into personalized decisions~\cite{zhang2020multimodal,liu2024mmrec}. 
Among these modalities, vision and language are particularly powerful for creating rich item representations~\cite{wei2019mmgcn,yu2023multi,zhou2023comprehensive,ren2024advances,malitesta2025formalizing}. 
We conceptualize the specific task of leveraging them for recommendation as Vision-Language-Recommendation (VLR). 
VLR models aim to move beyond simple interaction data by interpreting the actual content of items, which can alleviate data sparsity and capture a more nuanced understanding of user preferences~\cite{liu2019user,ren2024advances,wei2024towards,yu2025mind,zhou2025large}.

The typical approach in a centralized setting involves a two-stage process. 
First, powerful pre-trained foundation models are used to extract high-level semantic features from item images and textual descriptions~\cite{zhang2021unbert,hou2022towards,bian2023multi,geng2023vip5,sun2023universal,wang2023missrec,lu2023scaling,fu2024exploring,pan2024federated,zhou2025large}.
Second, a dedicated fusion module combines these unimodal representations into a single comprehensive embedding for each item~\cite{radford2021learning,zhang2021unbert,deng2024unlocking,liu2024multimodal,wei2024towards}. 
This fused representation is then used to predict user preferences. 
While effective, these centralized models require unrestricted access to all user and item data, which motivates our work to adapt these sophisticated VLR techniques to a privacy-preserving environment where data remains decentralized on devices. \looseness=-1

\subsection{Federated Learning for On-Device Personalization}
On-device recommendation involves training models on a user's local device to preserve privacy~\cite{li2024federated,yin2024device}. 
Federated Learning (FL)~\cite{mcmahan2017communication} provides the foundational framework, enabling collaborative training without centralizing sensitive data. 
In the context of recommendation, this creates a one-user-per-client setting for personalization~\cite{zhang2023dual,zhang2023graph,zhang2023lightfr}.
Incorporating rich item content into this setting introduces a key personalization challenge. 
Users weigh modalities like vision and text differently when making choices~\cite{liu2019user,niu2021review,lei2023learning}. 
This diversity makes a single global modal fusion module across all users suboptimal, as a fixed rule for all users fails to capture their nuanced individual tastes for different modality~\cite{zhang2024federateda,yuan2024hetefedrec,kong_oh-fedrec_2025}. 

Most on-device recommendation models are content-agnostic~\cite{ammad2019federated,lin2020fedrec,liang2021fedrec++,luo2022personalized,perifanis2022federated,zhang2023dual,zhang2023graph,zhang2025multifaceted}, relying solely on user-item interaction IDs. 
While these methods address statistical heterogeneity from non-IID data~\cite{allouah2023robust,li2024federated,yuan2024hetefedrec}, their dependence on IDs prevents the integration of rich item content. 
A few recent studies have incorporated multiple modalities of content~\cite{li2024towards,feng2024robust}, but they typically impose a uniform fusion logic on all users. 
Therefore, developing a user-oriented fusion module that can be personalized on each user's device remains a critical open problem in content-aware on-device recommendation.

\section{Problem Formulation}
\label{sec:problem_formulation}

We consider the user-oriented federated setting for on-device VLR tasks, where the system consists of a set of users \(\mathcal{U}\) and a set of items \(\mathcal{I}\). 
Each user \(u \in \mathcal{U}\) acts as a distinct client, holding their private interaction data. 
We represent the historical interactions as a binary matrix \(\mathbf{R} \in \{0, 1\}^{|\mathcal{U}| \times |\mathcal{I}|}\), where \(r_{u,i} = 1\) signifies that user \(u\) has interacted with item \(i\). 
The set of observed interactions for user \(u\) is \(\mathcal{O}_u = \{i \in \mathcal{I} \mid r_{u,i} = 1\}\).
Each item \(i \in \mathcal{I}\) is described by features from a set of modalities \(\mathcal{M}\).
For a given modality \(m \in \mathcal{M}\), the features for all items are represented by a matrix \(\mathbf{E}_{m} \in \mathbb{R}^{|\mathcal{I}| \times d_m}\), where \(\mathbf{e}_{i,m} \in \mathbb{R}^{d_m}\) is the feature vector for item \(i\). 
Visual \((v)\) and textual \((c)\) modalities all are stored on the server to ensure client-side efficiency. 

\begin{figure}[!t]
    \centering
    \includegraphics[width=1\linewidth]{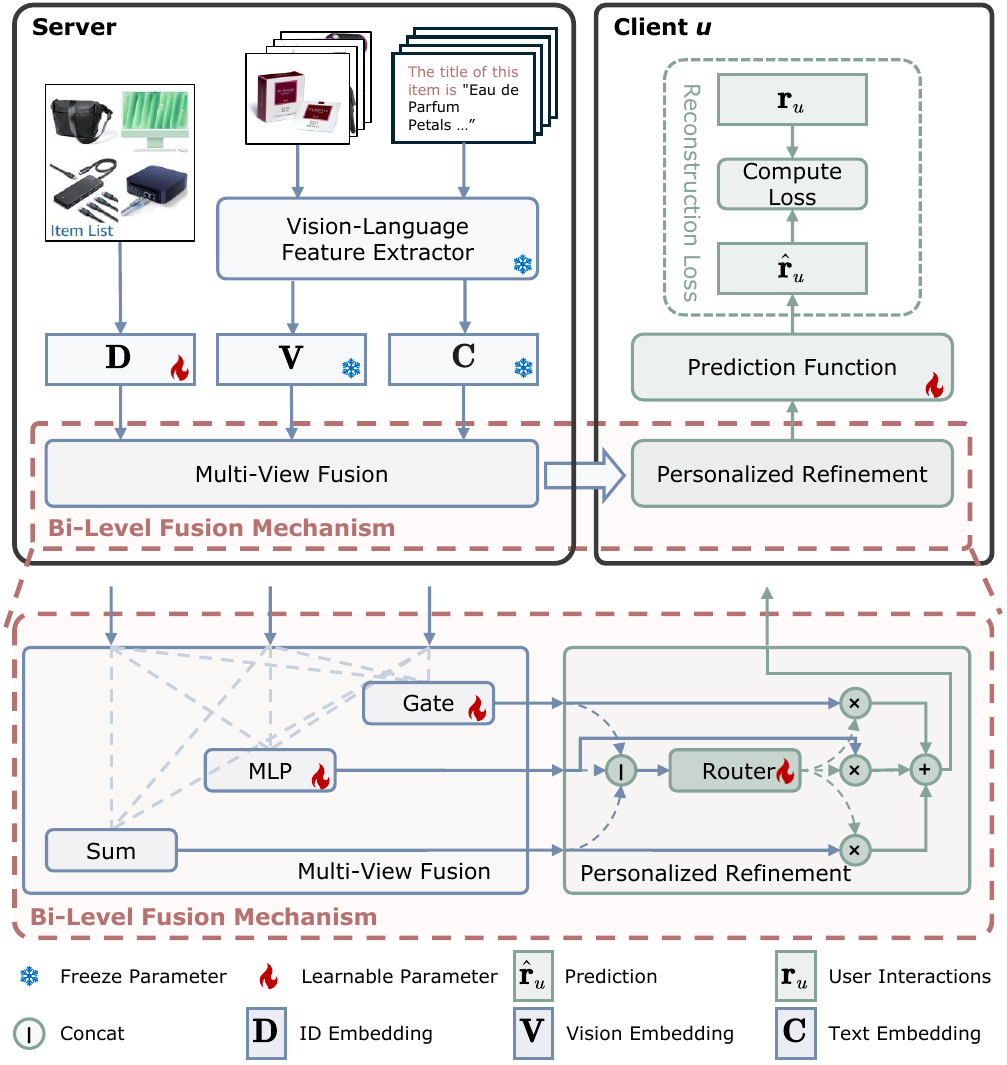}
    \caption{
        The framework of FedVLR.
        It comprises two components: 
        (1) \textbf{Server-Side Multi-View Fusion}, which generates diverse pre-fused feature views from visual-language content,
        and (2) \textbf{Client-Side Personalized Refinement}, which dynamically combines these views based on the user's private interaction history for on-device VLR.
    }
    \label{fig:framework}
\end{figure}

\section{Methodology}
\label{sec:methodology}

This section presents our proposed FedVLR framework, with its overall architecture depicted in Fig.~\ref{fig:framework}. 
The core of FedVLR lies in a novel Bi-Level Fusion Mechanism that learns across the server and clients to achieve on-device VLR within the user-oriented federated framework.

\subsection{Preliminary}

As depicted on the server-side of Fig.~\ref{fig:framework}, FedVLR's initial step is to prepare a comprehensive set of item representations. 
To avoid burdening user devices, this process is handled entirely by the server. 
A frozen pre-trained VLM \(h(\cdot)\) is employed in a one-time process to transform raw visual features \(\mathbf{E}_v\) and textual features \(\mathbf{E}_c\) into high-level semantic embeddings, yielding the vision and text embedding \(\mathbf{V}\) and \(\mathbf{C}\): \looseness=-1
\begin{equation}
  \mathbf{V} = h(\mathbf{E}_v),
  \quad
  \mathbf{C} = h(\mathbf{E}_c)\,.
  \label{eq:encode}
\end{equation}
Alongside these static content embeddings, FedVLR also maintains a globally learnable ID embedding \(\mathbf{D} \in \mathbb{R}^{|\mathcal{I}| \times d}\), designed to capture collaborative signals from user-item interactions. 
Collectively, the set \(\{\mathbf{D}, \mathbf{V}, \mathbf{C}\}\) provides a multi-view representation for all items available to the system.

\subsection{Bi-Level Fusion Mechanism}

To address both statistical and preference heterogeneity, FedVLR introduces the Bi-Level Fusion Mechanism (BLFM), which learns at two coordinated levels: server-side multi-view fusion and client-side personalized refinement.

\noindent \textbf{Server-Side Multi-View Fusion}.
Instead of modeling a single unified item representation, the server generates a diverse set of fused feature views by employing a collection of distinct learnable fusion operators \(\mathcal{G} = \{g_j\}_{j=1}^{|\mathcal{G}|}\).
Each operator \(g_j\) takes the full set of item representations \(\{\mathbf{V}, \mathbf{C}, \mathbf{D}\}\) as input to produce a unique fused view \(\mathbf{F}_j\):
\begin{equation}
  \mathbf{F}_j = g_j\bigl(\mathbf{V}, \mathbf{C}, \mathbf{D};\,\gamma_j\bigr)\,
  \label{eq:server_fuse}
\end{equation}
where \(\gamma_j\) are the learnable parameters of the \(j\)-th fusion operator. 
As shown in the lower panel of Fig.~\ref{fig:framework}, these operators can include simple strategies like element-wise \emph{Sum}, as well as more complex parameterized functions like a multi-layer perceptron \emph{MLP} or a \emph{Gate} mechanism. 
The fused views \(\{\mathbf{F}_j\}\) are then broadcast to the client devices.

\noindent \textbf{Client-Side Personalized Refinement}.
Upon receiving the set of pre-fused views, each client \(u\) performs personalized refinement via a MoE module, allowing FedVLR to adapt to the individual user's preferences. 
As shown in Fig.~\ref{fig:framework}, the client employs a local lightweight \emph{Router} module \(\phi_u\), parameterized by the local parameters \(\varphi_u\). 
It takes the server-provided item views \(\{\mathbf{F}_j\}\) and the user's history \(\mathbf{r}_u\) as input to dynamically compute a set of importance weights \(\mathbf{w}_u\):
\begin{equation}
  \mathbf{w}_u = \mathrm{softmax} \left(\phi_u\left(\{\mathbf{F}_j\}_{j=1}^{|\mathcal{G}|},\,\mathbf{r}_u;\,\varphi_u\right)\right).
  \label{eq:router}
\end{equation}
Each weight \(w_{u,j}\) in the vector \(\mathbf{w}_u \in \mathbb{R}^{|\mathcal{G}|}\) quantifies how much user \(u\) values the \(j\)-th feature view. 
The client then models its final personalized item representation \(\bar{\mathbf{F}}^{(u)}\) by computing a weighted sum of the global views:
\begin{equation}
  \bar{\mathbf{F}}^{(u)}
  = \sum_{j=1}^{|\mathcal{G}|} w_{u,j}\,\mathbf{F}_j.
  \label{eq:client_fuse}
\end{equation}

\subsection{Objective Function}

With the personalized item representation \(\bar{\mathbf{F}}^{(u)}\) for user \(u\), the preference score \(\hat{r}_{ui}\) for an item \(i\) is computed by a local prediction function \(f\), parameterized by local parameters \(\theta_u\):
\begin{equation}
  \hat{r}_{ui} = f(\bar{\mathbf{F}}_i^{(u)};\,\theta_u).
  \label{eq:pred}
\end{equation} 
The training process aims to find the optimal set of parameters \(\mathbf{\Theta} = (\{\mathbf{D}, \{\gamma_j\}\}, \{\theta_u, \varphi_u\})\) for minimizing the reconstruction loss between the predicted scores \(\hat{r}_{ui}\) and the user interactions \(r_{ui}\), aggregated across all users. 
The joint optimization problem of FedVLR is then formulated as follows:
\begin{equation}\label{eq:recon}
\begin{aligned}
  \min_{\Theta}\quad
    &\mathcal{J}(\Theta)
     = \sum_{u\in\mathcal{U}} \alpha_u\,\mathcal{J}_u(\Theta),\\
  \text{s.t.}\quad
    &\begin{alignedat}{2}
      \mathbb{E}_\xi\bigl[\|g_u(\Theta)-\nabla\mathcal{J}_u(\Theta)\|^2\bigr]
        &\le \sigma^2,\\
      \mathbb{E}_u\bigl[\|\nabla\mathcal{J}_u(\Theta)
        - \nabla\mathcal{J}(\Theta)\|^2\bigr]
        &\le \zeta^2.
    \end{alignedat}
\end{aligned}
\end{equation}
In Eq.~\eqref{eq:recon}, we assume that each client's objective, denoted by \(\mathcal{J}_u (\mathbf{\Theta}) = \mathcal{L}_{u}(\hat{r}_{ui},\,r_{ui})\), belongs to the class of L-smooth functions \(\mathcal{F}_L\). 
Furthermore, the optimization learns under the standard conditions of bounded variance \(\sigma^2\) for stochastic gradients on each client  and \(\zeta^2\) for the gradient dissimilarity across the population of clients, which are standard prerequisites for establishing convergence in federated training. 
The procedure for optimizing this objective is detailed in Alg.~\ref{alg:algorithm}.

\begin{algorithm}[!t]
    \small
    \begin{minipage}{1\linewidth}
    \caption{The training algorithm for FedVLR}
    \label{alg:algorithm}
    
    \textbf{Input}: User Interaction History \(\mathbf{R}\), Learning Rate \(\eta\), Number of Communication Round \(T\), Number of Local Training Epoch \(A\) \\
    \textbf{Initialize}: User Embeddings \(\{\theta_u\}\), User-specific Parameters \(\{\varphi_u\}\), Strategy Parameters \(\{\gamma_j\}\), ID Embeddings \(\mathbf{D}\) \\
    \textbf{MultiStrategyFusion}: 
    \begin{algorithmic}[1] 
        \State \(\mathbf{V} \leftarrow h(\mathbf{E}_v)\), \(\mathbf{C} \leftarrow h(\mathbf{E}_c)\) according to Eq.~\eqref{eq:encode};

        \For{\(t=1,2,\dots,T\)}
            \State \(S_t \leftarrow\) randomly select \(n_s\) from \(n\) clients;
            \State Compute \(\mathbf{F}_j\) according to Eq.~\eqref{eq:server_fuse}, \(j = 1, 2, \cdots |\mathcal{G}|\);
            
            \ForAll{client index \(u \in S_t\)};
                \State \(\hat{\mathbf{r}}_u\), \(\nabla_{\mathbf{D}}^{(u)}\), \(\{\nabla^{(u)}_{j}\gamma\} \leftarrow \text{ClientUpdate}(\{\mathbf{F}_j\}_{j=1}^{|\mathcal{G}|})\);
            \EndFor

            \State \(\mathbf{D} \leftarrow \mathbf{D} - \eta \sum_{u \in S_t} \alpha_u \nabla_{\mathbf{D}}^{(u)}\);
            \State \(\gamma_j \leftarrow \gamma_j - \eta \sum_{u \in S_t} \alpha_u \nabla^{(u)}_{j}\gamma\), \(j = 1,2,\cdots,|\mathcal{G}|\);
        \EndFor
        \State \textbf{return}: \(\hat{\mathbf{R}} = [\hat{r}_1, \hat{r}_2, \dots, \hat{r}_n]^T\)

    \end{algorithmic}

    \textbf{PersonalizedRefinement}: 
    \begin{algorithmic}[1] 
        \State \(\nabla_{\mathbf{D}}^{(u)} \leftarrow 0\);
        \State \(\nabla^{(u)}_{j}{\gamma} \leftarrow 0\), \(j = 1, 2, \cdots |\mathcal{G}|\);
        \For{\(a=1,2,\dots,A\)}  
            \State Compute \(\nabla_{\theta_u}\), \(\nabla_{\varphi_u}\), \(\nabla_{\mathbf{D}}\) and \(\{\nabla_{\gamma_j}\}\) according to Eq.~\eqref{eq:recon};
            \State Update local prediction parameters \(\theta_u \leftarrow \theta_u - \eta \nabla_{\theta_u}\);
            \State Update local router parameters \(\varphi_u \leftarrow \varphi_u - \eta \nabla_{\varphi_u}\);
            \State Accumulate gradients \(\nabla_{\mathbf{D}}^{(u)} \leftarrow \nabla_{\mathbf{D}}^{(u)} + \nabla_{\mathbf{D}}\);
            \State Accumulate \(\nabla^{(u)}_{j}{\gamma} \leftarrow \nabla^{(u)}_{j}{\gamma} + \nabla_{\gamma_j}\), \(j = 1, 2, \cdots |\mathcal{G}|\);
            \State Compute \(\hat{\mathbf{r}}_u\) according to Eq.~\eqref{eq:pred};
        \EndFor

        \State \textbf{return}: \(\hat{\mathbf{r}}_u\), \(\nabla_{\mathbf{D}}^{(u)}\), \(\{\nabla^{(u)}_{j}{\gamma}\}\)
    \end{algorithmic}
    \end{minipage}
\end{algorithm}

\section{Theoretical Analysis}

\subsection{Convergence Analysis}

In each communication round \(t\), clients perform \(A\) local updates, which inevitably introduces a drift between the locally trained parameters and the global model state. 
Despite this drift, by building on the standard assumptions for the objective function in Eq.~\eqref{eq:recon}, we can guarantee that our proposed FedVLR converges to a stationary point of $\mathcal{J}$ in a manner consistent with rates of typical non-convex FL problems:

\begin{theorem}[Convergence of FedVLR]
\label{thm:convergence}
Let the number of participating clients per round be $n_s$. 
After $T$ communication rounds with learning rate $\eta$, the algorithm satisfies:
\begin{equation}
\small
\frac{1}{T}\sum_{t=0}^{T-1} \mathbb{E}\bigl[\|\nabla \mathcal{J}(\mathbf{\Theta}^t)\|^2\bigr]
\le O\Bigl(\frac{1}{\sqrt{T}}\Bigr)
+ O\Bigl(\frac{A}{T}\Bigr)
+ O\Bigl(\frac{\zeta^2}{n_s}\Bigr).
\end{equation}
\end{theorem}

Theorem~\ref{thm:convergence} demonstrates that FedVLR is theoretically sound, achieving a standard federated convergence rate, validating that our personalized fusion architecture can effectively address preference heterogeneity without impeding convergence. 
The detailed proof is provided in the appendix.

\subsection{Complexity Analysis}

Here, we analyze the computational and storage costs of FedVLR. 
The server stores the global item embeddings, including the ID embeddings $\mathbf{D}$ and the pre-computed multimodal embeddings $\mathbf{V}$ and $\mathbf{C}$, requiring $O(|\mathcal{I}|d)$ space, where $d$ is the embedding dimension. 
In each communication round, the server's main computational cost comes from generating the $|\mathcal{G}|$ fused views, resulting in a time complexity of $O(|\mathcal{G}||\mathcal{I}|d)$.
Each client stores its local model parameters $K$. 
The primary storage overhead is from receiving the $|\mathcal{G}|$ feature views, requiring $O(|\mathcal{G}||\mathcal{I}|d + K)$ space temporarily during training. 
The client's per-iteration time complexity for local updates, involving the prediction function \(f\) and the BLFM router \(\phi_u\), is denoted as \(O(P)\). 
Therefore, the total system space complexity is roughly \(O(|\mathcal{U}| (|\mathcal{G}| |\mathcal{I}| d + K) + |\mathcal{I}| d)\), and the total time complexity per round involves server computation plus aggregated client computation, scaling approximately as \(O(|\mathcal{G}| |\mathcal{I}| d + n_s A P)\) after performing $A$ local update steps. 
Therefore, FedVLR adds a manageable overhead that is well-justified by its ability to model complex, personalized multimodal preferences.

\subsection{Privacy Preservation}
\label{sec:privacy}

FedVLR ensures privacy based on the foundational FL principle of data localization, where raw user interactions \(\mathbf{r}_u\) remain on-device. 
Crucially, all personalization components, including the user-specific router \(\phi_u\) and its parameters \(\varphi_u\), are kept entirely local. 
This design ensures individual modality preferences are never exposed. 
The gradients transmitted to the server are structurally analogous to those in standard ID-based frameworks \cite{li2024federated,zhang2023dual,li2024personalized,feng2024robust,wu2024towards}, thus introducing no new attack surfaces. 
Furthermore, FedVLR is compatible with advanced privacy-enhancing technologies, which our experiments show can be incorporated with an acceptable performance trade-off.


\section{Experiments}

\subsection{Datasets}

We conduct a comprehensive evaluation on seven public datasets spanning three distinct domains shown in Table \ref{table:datasets}:
\begin{itemize}
\item \textbf{Amazon Review Datasets}\footnote{https://amazon-reviews-2023.github.io/}~\cite{hou2024bridging}: 
Two review datasets are selected for the e-commerce domain, i.e., All\_Beauty (Beauty) and Gift\_Cards (Cards).
\item \textbf{MovieLens-latest-small (ML)}\footnote{https://grouplens.org/datasets/movielens/latest/}~\cite{harper2015movielens}: 
A classic dataset for the movie recommendation.
\item \textbf{NineRec Datasets}\footnote{https://github.com/westlake-repl/NineRec}~\cite{zhang2024ninerec}: 
Four datasets from the short-video domain are selected: KU, Bili\_Food (Food), Bili\_Dance (Dance), and Bili\_Movie (Movie).
\end{itemize}
This diverse selection across different domains, platforms, and scales provides a robust testbed for assessing FedVLR's adaptability and generalizability under realistic and extreme sparse data conditions (over 95\%) in the user-oriented federated settings. 
Following standard practice, we filter out users with fewer than 5 interactions to mitigate cold-start issues. 
For data preprocessing, missing images are imputed using the average visual features of existing items, and missing titles are replaced with the placeholder ``The title is missing."


\begin{table}[!tb]
\centering
{\small
\begin{tabular}{lcccc}
\hline
Dataset & \#Users & \#Items & \#Ratings & Sparsity \\ \hline
Beauty  & 253     & 356     & 2,535      & 97.19\%  \\
Cards   & 377     & 129     & 2,429      & 95.01\%  \\
ML      & 610     & 3,650    & 90,274     & 95.95\%  \\
KU      & 204     & 560     & 3,488      & 96.95\%  \\
Dance   & 10,231   & 1,676    & 80,086     & 99.53\%  \\
Food    & 5,990    & 1,125    & 36,482     & 99.46\%  \\
Movie   & 15,908   & 2,528    & 111,091    & 99.72\%  \\ \hline
\end{tabular}
}
\caption{
The statistical information of the datasets used in our work includes:
\#Interactions represents the number of observed interactions.
\#Users denotes the number of users.
\#Items indicates the number of items.
Sparsity is the percentage of \#Interactions out of the total possible ratings.
}
\label{table:datasets}
\end{table}

\subsection{Experimental Setup}

We evaluate FedVLR against two categories: (1) centralized multimodal models and (2) ID-based federated frameworks.

\textbf{Centralized VLR Baselines} serve as performance references with full data access in a non-federated environment:
\begin{itemize}
\item \textbf{VBPR}~\cite{he2016vbpr}: A Bayesian personalized ranking method incorporating visual features;
\item \textbf{BM3}~\cite{zhou2023bootstrap}: A self-supervised recommendation framework for multimodal data integration;
\item \textbf{MGCN}~\cite{yu2023multi}: A model that employs multi-view graph convolutions to fuse item's modal features;
\item \textbf{PGL}~\cite{yu2025mind}: Mines local graph structures of the user-item interaction to enhance performance.
\end{itemize}

\textbf{ID-based Federated Frameworks} represent diverse architectures for personalization in federated settings and are used as the backbone to test our approach\footnote{We do not include direct comparisons with AMMFRS \cite{feng2024robust} and FedMMR \cite{li2024towards} as their official code implementations were not publicly available at the time of our experiments, preventing a fair and reproducible comparison.}. 
For each baseline, we compare its original performance with its performance when enhanced by our framework (``+ours"):
\begin{itemize}
\item \textbf{FedAvg}~\cite{mcmahan2017communication}: The foundational federated algorithm based on weighted model averaging;
\item \textbf{FCF}~\cite{ammad2019federated}: Federated collaborative filtering using equal-weighted averaging;
\item \textbf{FedNCF}~\cite{perifanis2022federated}: A framework with local user embeddings and global item embeddings;
\item \textbf{PFedRec}~\cite{zhang2023dual}: Enables two-way personalization through server-side aggregation;
\item \textbf{FedRAP}~\cite{li2024federated}: Uses additive modeling for fine-grained user and item personalization.
\end{itemize}

To further validate FedVLR's design, we also include two additional variants: 
a naive federated adaptation of VBPR using FedAvg (``+VBPR"), and an architectural ablation that replaces our BLFM module with standard MLPs  (``+MLP").

\textbf{Evaluation Protocol.} 
To ensure a fair comparison, all methods are evaluated under the same protocol. 
We treat observed interactions as positive samples and all other items as negative samples. 
We then evaluate performance by ranking the predicted scores for all candidate items in descending order, after masking items seen during training. 
We use two standard metrics for implicit feedback \cite{he2020lightgcn}: Hit Rate (HR@K) and Normalized Discounted Cumulative Gain (NDCG@K), setting K=50 for all experiments. 
The rank-sensitive nature of NDCG is particularly crucial for our evaluation, as it directly measures the quality of fine-grained personalization that our multimodal fusion aims to achieve.

\subsection{Experimental Setting}

Following prior work \cite{zhang2023dual,li2024federated}, we employ negative sampling during training and use the leave-one-out strategy for validation and testing. 
Hyperparameters for all baselines are tuned via grid search on the validation set, including the learning rate \(\eta\) within \(\{10^i \mid i = -4, \ldots, -1\}\) and method-specific parameters. 
We use CLIP \cite{radford2021learning} as the foundation model for extracting visual and textual features, implemented via OpenCLIP \cite{ilharco_gabriel_2021_5143773} with pre-trained ViT-B-32 weights from OpenAI \cite{Radford2021LearningTV,schuhmann2022laionb}. 
A linear mapping layer projects multimodal and ID embeddings to a shared latent dimension of 64 for efficiency. 
The training batch size is 2048. 
All models incorporating hidden layers use a three-layer MLP structure. 
For federated methods, clients perform \(5\) local training epochs per communication round using a consistent aggregation strategy for both vanilla and FedVLR-integrated versions.

\begin{table*}[t]
\centering
{\setlength{\tabcolsep}{1.6mm}\small
\begin{tabular}{lcccccccccccccc}
\hline
\multicolumn{1}{c}{}                         & \multicolumn{2}{c}{\textbf{Beauty}}      & \multicolumn{2}{c}{\textbf{Cards}}       & \multicolumn{2}{c}{\textbf{ML}}  & \multicolumn{2}{c}{\textbf{KU}}          & \multicolumn{2}{c}{\textbf{Dance}}                           & \multicolumn{2}{c}{\textbf{Food}}       & \multicolumn{2}{c}{\textbf{Movie}}                           \\
\multicolumn{1}{c}{\multirow{-2}{*}{Method}} & HR             & NDCG           & HR             & NDCG           & HR             & NDCG          & HR             & NDCG           & HR                       & NDCG                     & HR             & NDCG          & HR                       & NDCG                     \\ \hline
VBPR                                         & 24.11          & 6.65           & 73.74          & 29.34          & 23.11          & 6.98          & 44.12          & 15.47          & 23.06                    & 7.47                     & 24.72          & 7.90          & 14.29                    & 4.86                     \\
BM3                                          & 18.58          & 4.15           & 83.55          & 34.46          & 18.36          & 4.91          & 41.67          & 15.95          & 26.54                    & 8.56                     & 25.63          & 8.20          & 19.17                    & 6.66                     \\
MGCN                                         & 29.25          & 6.94           & 80.11          & 33.27          & 25.74          & 7.10          & 33.82          & 15.08          & 23.63                    & 7.67                     & 24.09          & 7.77          & 16.97                    & 5.74                     \\
PGL                                          & 22.53          & 6.60           & 81.17          & 32.18          & 26.07          & 8.37          & 46.57          & 20.13          & 24.83                    & 8.25                     & 36.16          & 13.56         & 17.37                    & 5.94                     \\ \hline
FedAvg                                       & 17.00          & 3.82           & 64.19          & 23.20          & 11.80          & 3.79          & 11.27          & 5.62           & 7.82                     & 2.01                     & 6.96           & 1.91          & 5.52                     & 1.58                     \\
\hspace{0.5em}+ VBPR          & 18.18          & 4.42           & 64.99          & 19.03          & 7.87           & 3.36          & 51.47          & 14.91          & \textbackslash{} & \textbackslash{} & 9.82           & 2.91          & \textbackslash{} & \textbackslash{} \\
\hspace{0.5em}+ MLP           & 18.58          & 5.39           & 68.17          & 23.20          & 7.87           & 3.28          & 39.22          & 12.40          & 8.94                     & 2.18                     & 9.63           & 2.70          & 5.63                     & 1.60                     \\
\hspace{0.5em}+ ours          & 19.37          & 5.51           & 70.29          & 27.60          & 12.30          & 3.81          & 54.90          & \textbf{16.04} & \textbf{10.07}           & \textbf{2.61}            & 9.48           & 2.68          & 5.64                     & 1.60                     \\[1pt]
FCF                                          & 28.46          & 8.87           & 69.50          & 25.16          & 11.64          & 3.72          & 22.55          & 6.43           & 6.76                     & 1.77                     & 8.20           & 2.19          & 5.81                     & 1.63                     \\
\hspace{0.5em}+ MLP           & \textbf{41.11} & 11.77          & 71.35          & 20.27          & 11.64          & 3.72          & 36.76          & 12.40          & 7.26                     & 1.74                     & 8.99           & 2.27          & 6.28                     & 1.73                     \\
\hspace{0.5em}+ ours          & 37.15          & \textbf{12.04} & 72.68          & 28.51          & \textbf{13.28} & 3.91          & \textbf{55.39} & 15.97          & 8.06                     & 2.17                     & 9.18           & 2.47          & \textbf{6.67}            & 1.80                     \\[1pt]
FedNCF                                       & 17.79          & 4.32           & 42.71          & 11.13          & 3.93           & 1.14          & 11.76          & 2.49           & 2.21                     & 0.55                     & 5.48           & 1.63          & 1.85                     & 0.54                     \\
\hspace{0.5em}+ MLP           & 20.16          & 5.07           & 70.82          & 22.73          & 10.82          & 3.54          & 28.43          & 6.91           & 6.01                     & 1.47                     & 9.18           & 2.40          & 3.70                     & 1.04                     \\
\hspace{0.5em}+ ours          & 21.74          & 5.08           & 72.68          & 28.38          & 11.31          & 3.01          & 43.63          & 12.18          & 7.08                     & 1.87                     & 8.43           & 2.43          & 4.06                     & 1.25                     \\[1pt]
PFedRec                                      & 17.79          & 4.56           & 32.63          & 8.67           & 11.31          & 3.67          & 6.37           & 2.30           & 2.52                     & 0.68                     & 3.27           & 0.93          & 1.70                     & 0.42                     \\
\hspace{0.5em}+ MLP           & 21.34          & 4.67           & 69.76          & 25.31          & 11.97          & 3.38          & 39.22          & 8.52           & 8.52                     & 2.10                     & 8.68           & 2.50          & 5.41                     & 1.41                     \\
\hspace{0.5em}+ ours          & 24.90          & 5.72           & 70.29          & 27.76          & 12.30          & 3.81          & 45.59          & 15.05          & 7.46                     & 1.92                     & 10.10          & 2.55          & 5.72                     & 1.48                     \\[1pt]
FedRAP                                       & 23.72          & 9.18           & 62.60          & 22.29          & 11.64          & 3.76          & 47.06          & 10.86          & 6.82                     & 1.86                     & 6.96           & 1.91          & 4.53                     & 1.36                     \\
\hspace{0.5em}+ MLP           & 32.02          & 10.80          & 64.19          & 27.57          & 12.30          & 3.58          & 48.53          & 12.12          & 7.36                     & 1.76                     & 8.68           & 2.21          & 4.56                     & 1.42                     \\
\hspace{0.5em}+ ours          & 38.34          & 11.39          & \textbf{73.74} & \textbf{28.76} & 12.79          & \textbf{4.15} & 48.24          & 13.05          & 8.95                     & 2.47                     & \textbf{11.42} & \textbf{2.94} & 5.86                     & \textbf{1.99}            \\ \hline
\end{tabular}
}
\caption{
Performance comparison of our FedVLR (``+ours") against centralized and federated baselines, reporting HR@50 (HR) and NDCG@50 (NDCG) in percent. 
Federated baselines are enhanced with different fusion modules for ablation: 
``+MLP" uses a generic MLP-based fusion, while ``+VBPR" is a federated adaptation of VBPR. 
The highest federated result per column is in \textbf{bold}, and \textbackslash{} denotes runs that failed due to memory constraints. 
The results consistently demonstrate that our proposed FedVLR is a more effective and robust approach for enhancing personalized federated VLR frameworks than generic fusion alternatives.
}
\label{tab:main_results}
\end{table*}

\begin{table}[tb]
\centering
{\setlength{\tabcolsep}{1mm}\small
\begin{tabular}{lccccccc}
\hline
Model                               & \textbf{Beauty} & \textbf{Cards} & \textbf{ML} & \textbf{KU}  & \textbf{Dance} & \textbf{Food} & \textbf{Movie} \\ \hline
FedAvg                              & 22     & 8     & 233       & 287 & 107   & 72   & 161   \\
\hspace{0.5em}+ VBPR & 441    & 452   & 2,493      & 495 & 11,334 & 838  & 17,584 \\
\hspace{0.5em}+ MLP  & 68     & 24    & 700       & 107 & 321   & 216  & 485   \\
\hspace{0.5em}+ ours & 80     & 37    & 713       & 120 & 334   & 228  & 497   \\[1pt]
FCF                                 & 22     & 8     & 233       & 35  & 858   & 576  & 161   \\
\hspace{0.5em}+ MLP  & 68     & 24    & 700       & 107 & 321   & 216  & 485   \\
\hspace{0.5em}+ ours & 80     & 37    & 713       & 120 & 334   & 228  & 497   \\[1pt]
FedNCF                              & 88     & 75    & 556       & 108 & 1,535  & 921  & 1,182  \\
\hspace{0.5em}+ MLP  & 403    & 273   & 2,556      & 527 & 2,525  & 1,629 & 3,797  \\
\hspace{0.5em}+ ours & 440    & 311   & 2,594      & 564 & 2,561  & 1,666 & 3,833  \\[1pt]
PFedRec                             & 22     & 8     & 233       & 287 & 107   & 72   & 161  \\
\hspace{0.5em}+ MLP  & 68     & 24    & 700       & 107 & 321   & 216  & 485   \\
\hspace{0.5em}+ ours & 80     & 37    & 713       & 120 & 334   & 228  & 497   \\[1pt]
FedRAP                              & 45     & 16    & 467       & 71  & 214   & 144  & 323   \\
\hspace{0.5em}+ MLP  & 91     & 33    & 934       & 143 & 429   & 288  & 647   \\
\hspace{0.5em}+ ours & 103    & 45    & 946       & 155 & 441   & 300  & 659   \\ \hline
\end{tabular}
}
\caption{
Comparison of client-side parameters in thousands, detailing the parameter count for each ID-based federated baseline and its variants with different fusion modules. 
Notably, the parameter overhead of our FedVLR (``+ours'') is comparable to that of the generic MLP-based framework (``+MLP''), indicating the parameter efficiency of FedVLR.
}
\label{tab:parameter_count}
\end{table}

\subsection{Performance Analysis}

Tables \ref{tab:main_results} and \ref{tab:parameter_count} evaluate our proposed FedVLR’s performance and efficiency in the original federated frameworks and when enhanced with three different multimodal module variants: 
our personalized fusion framework (“+ours”), a generic MLP-based module for ablation (“+MLP”), and a federated adaptation of VBPR (“+VBPR”).


\noindent\textbf{Effectiveness of the Federated Architecture.}
The performance gains lies in FedVLR's ability to learn a personalized fusion model for each user. 
The generic MLP-based fusion provides a crucial comparison. 
As shown in Table~\ref{tab:parameter_count}, despite having a comparable parameter count, FedVLR consistently demonstrates superior performance, indicating the gain is a direct result of our BLFM architecture, not simply more parameters.
A notable exception is the HR performance on the small-scale Beauty dataset, where the simpler MLP-based fusion excels. 
This is likely because on datasets with very limited user data, a less adaptive model can be less prone to overfitting.
Nevertheless, the overall trend confirms that our client-side refinement successfully captures user-specific nuances that a one-size-fits-all mechanism cannot.
The direct federated adaptation of VBPR provides further evidence for our design. 
This naive adaptation proves impractical, as its excessive parameter count leads to out-of-memory failures on larger datasets and inferior performance where it runs. 
This underscores that an effective solution requires a purpose-built architecture, like our decoupled BLFM, designed fundamentally for the federated setting.

\noindent\textbf{Bridging the Gap to Centralized Performance.}
Our framework also demonstrates strong performance when compared to centralized models. 
Centralized methods learn from global interaction patterns, while our approach learns from isolated user data in a private decentralized manner. 
The performance dynamic between these two paradigms depends on the dataset's scale. 
On smaller datasets such as Beauty and KU, the global collaborative signal is sparse. 
In this regime, FedVLR's ability to learn a rich and personalized content model directly from user interactions becomes the decisive factor, often leading to superior performance. 
As the dataset scale and user population grow, the centralized model's access to a massive interaction matrix provides a strong advantage in discovering subtle and high-order collaborative patterns that are invisible to any single client. 
Even in these scenarios, FedVLR consistently narrows the performance gap. 
This key finding suggests that deep personalization from local data is a powerful alternative to learning from global signals, validating our privacy-preserving framework as a practical and effective approach.

\begin{figure*}[tb]
    \centering
    \begin{subfigure}[b]{0.245\linewidth}
        \includegraphics[width=\linewidth]{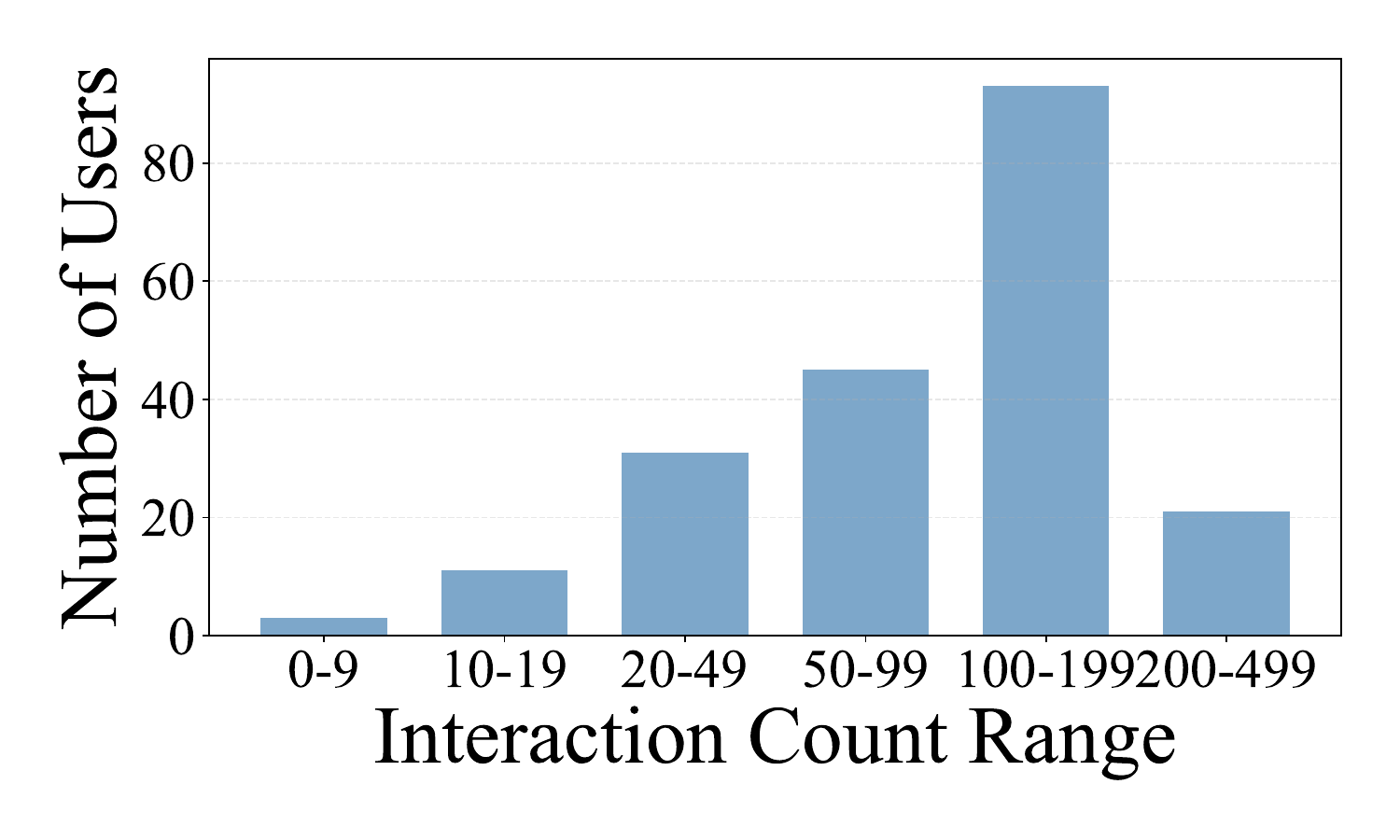}
        \caption{User Distribution}
        \label{fig:user_dist}
    \end{subfigure}
    \hfill
    \begin{subfigure}[b]{0.245\linewidth}
        \includegraphics[width=\linewidth]{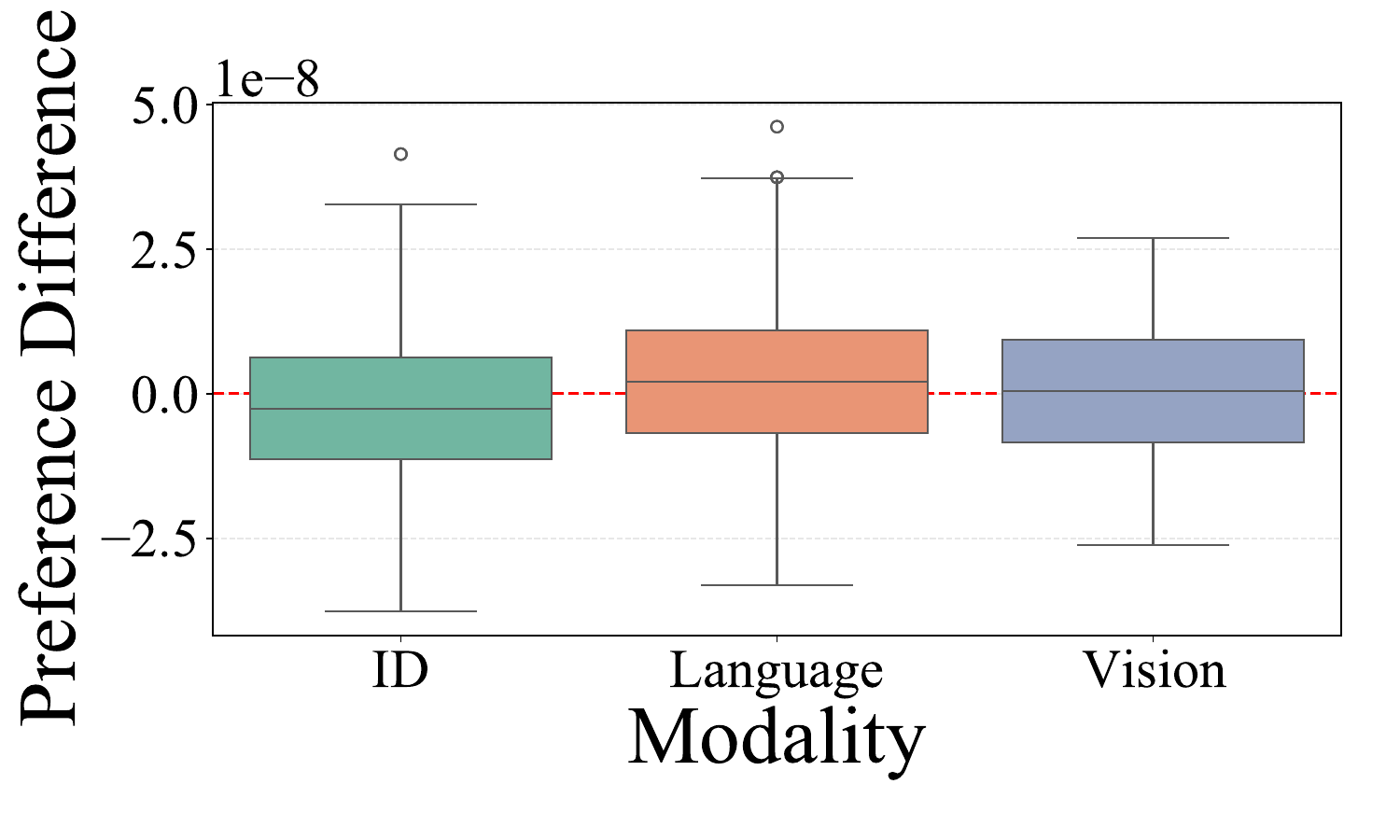}
        \caption{Modality Preferences}
        \label{fig:modal_heterogeneity}
    \end{subfigure}
    \hfill
    \begin{subfigure}[b]{0.245\linewidth}
        \includegraphics[width=\linewidth]{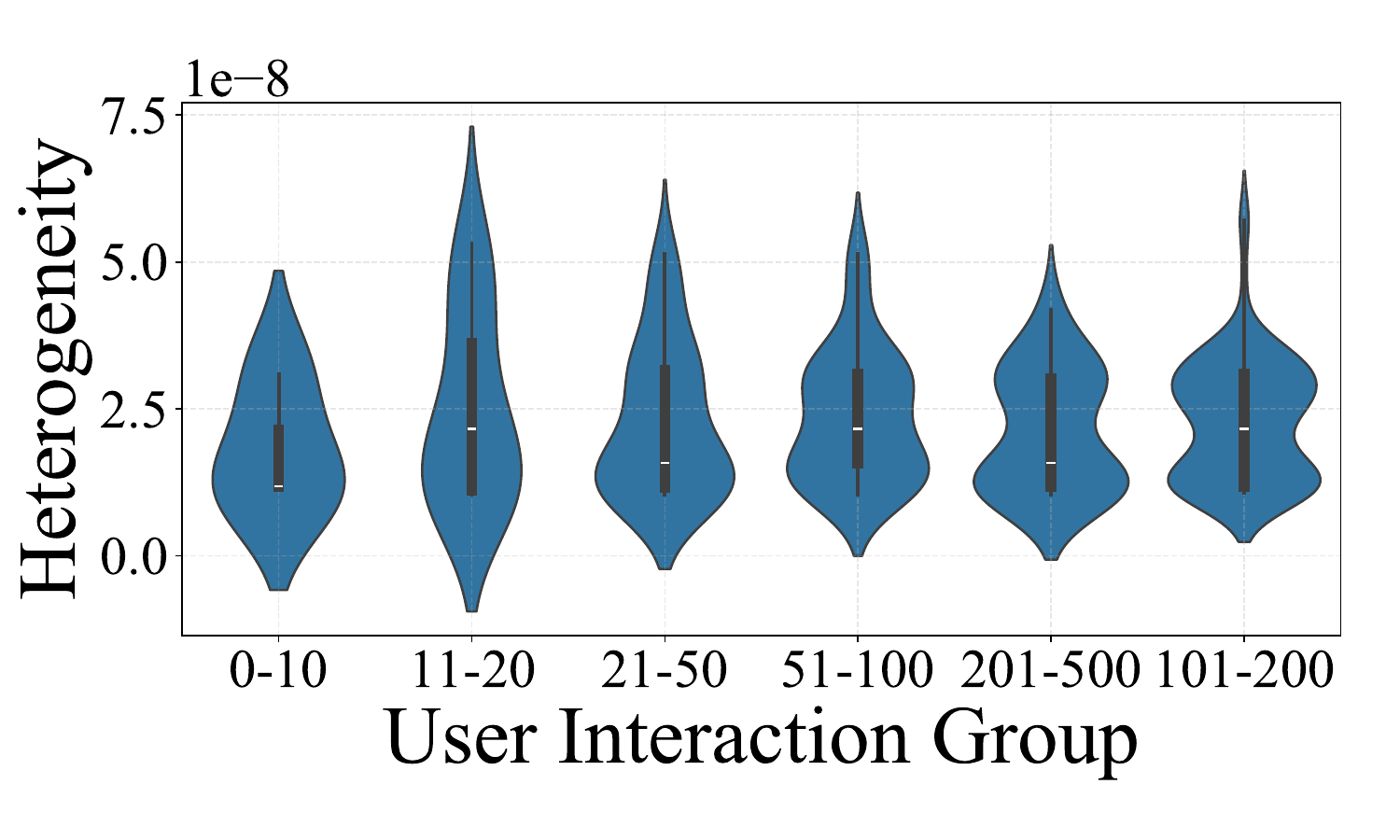}
        \caption{Heterogeneity Distribution}
        \label{fig:heterogeneity_dist}
    \end{subfigure}
    \hfill
    \begin{subfigure}[b]{0.245\linewidth}
        \includegraphics[width=\linewidth]{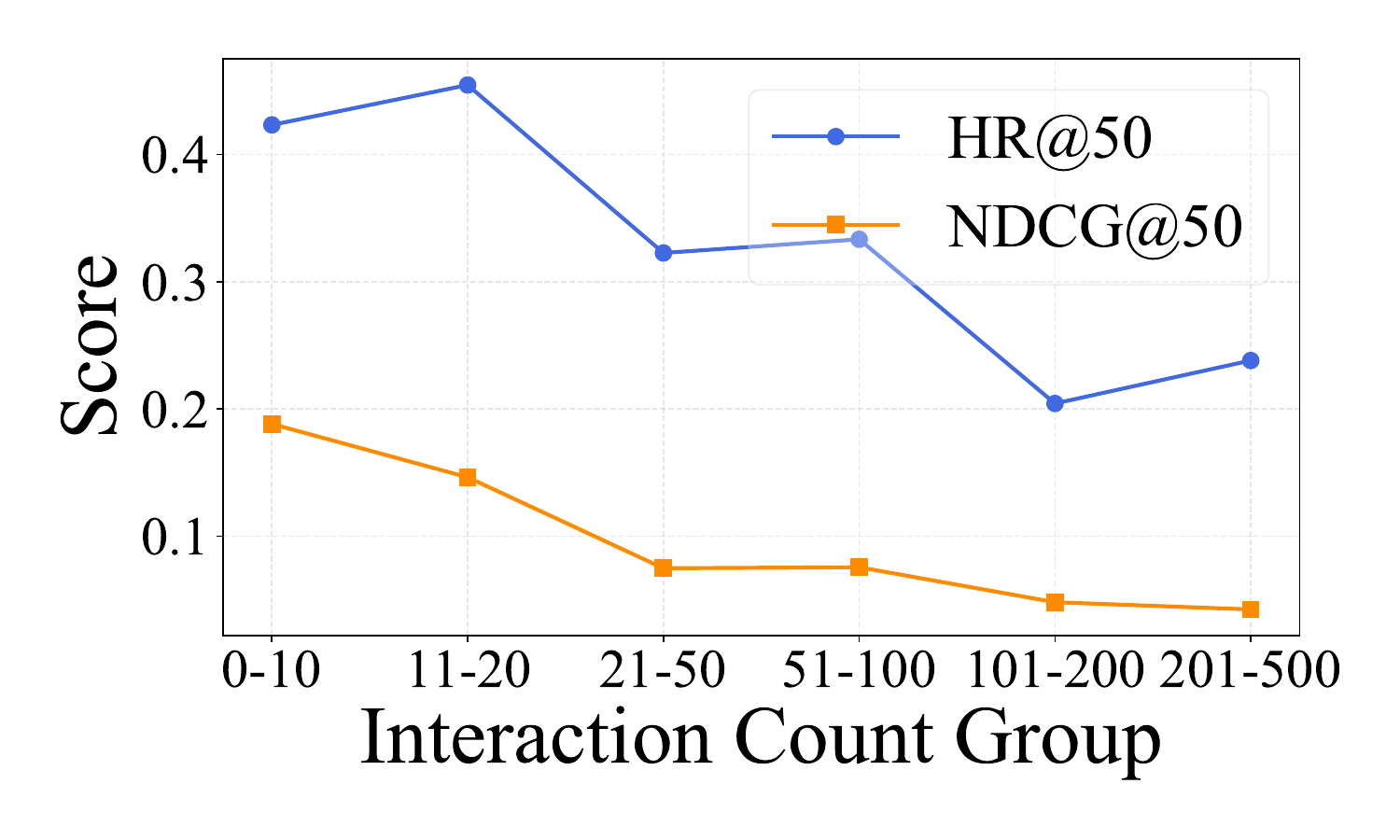}
        \caption{Performance Trend}
        \label{fig:perf_trend}
    \end{subfigure}
    \caption{
        Analysis of user characteristics learned by FedRAP enhanced with FedVLR on KU: 
        (a) User distribution by interaction count; 
        (b) User modality preference; 
        (c) User activity heterogeneity distribution; 
        (d) Performance trend across user groups.
    }
    \label{fig:analysis_composite}
\end{figure*}

\subsection{Ablation Studies}

We conduct a series of ablation studies to dissect the key components and properties of FedVLR. 
We analyze the impact of client heterogeneity, contribution of each modality, and the framework's robustness while privacy-enhancing.

\subsubsection{Analysis of Client Heterogeneity.}
We empirically investigate client heterogeneity learned by FedRAP enhanced with our FedVLR on the KU dataset. 
Fig.~\ref{fig:analysis_composite} reveals several key findings. 
Fig. \ref{fig:analysis_composite}a shows that users not only exhibit diverse activity levels, but they also show significant variance in their preferences for different modalities, especially for visual and textual content in Fig. \ref{fig:analysis_composite}b. 
Furthermore, this preference heterogeneity tends to increase for more active user groups, as indicated in Fig. \ref{fig:analysis_composite}.
This rising heterogeneity directly impacts model performance. 
As seen in Fig. \ref{fig:analysis_composite}d, accuracy tends to decrease for the most active and thus most heterogeneous user groups. 
Capturing the diverse and nuanced preferences of these users is a significant challenge for models with non-adaptive fusion logic, providing strong empirical validation for the central problem our paper addresses. 
FedVLR's client-side refinement module is designed specifically to tackle this preference heterogeneity, enabling the model to tailor its representations to each user.

\subsubsection{Contribution of Different Modalities.}

\begin{figure}[!tb]
    \centering
    \includegraphics[width=.95\linewidth]{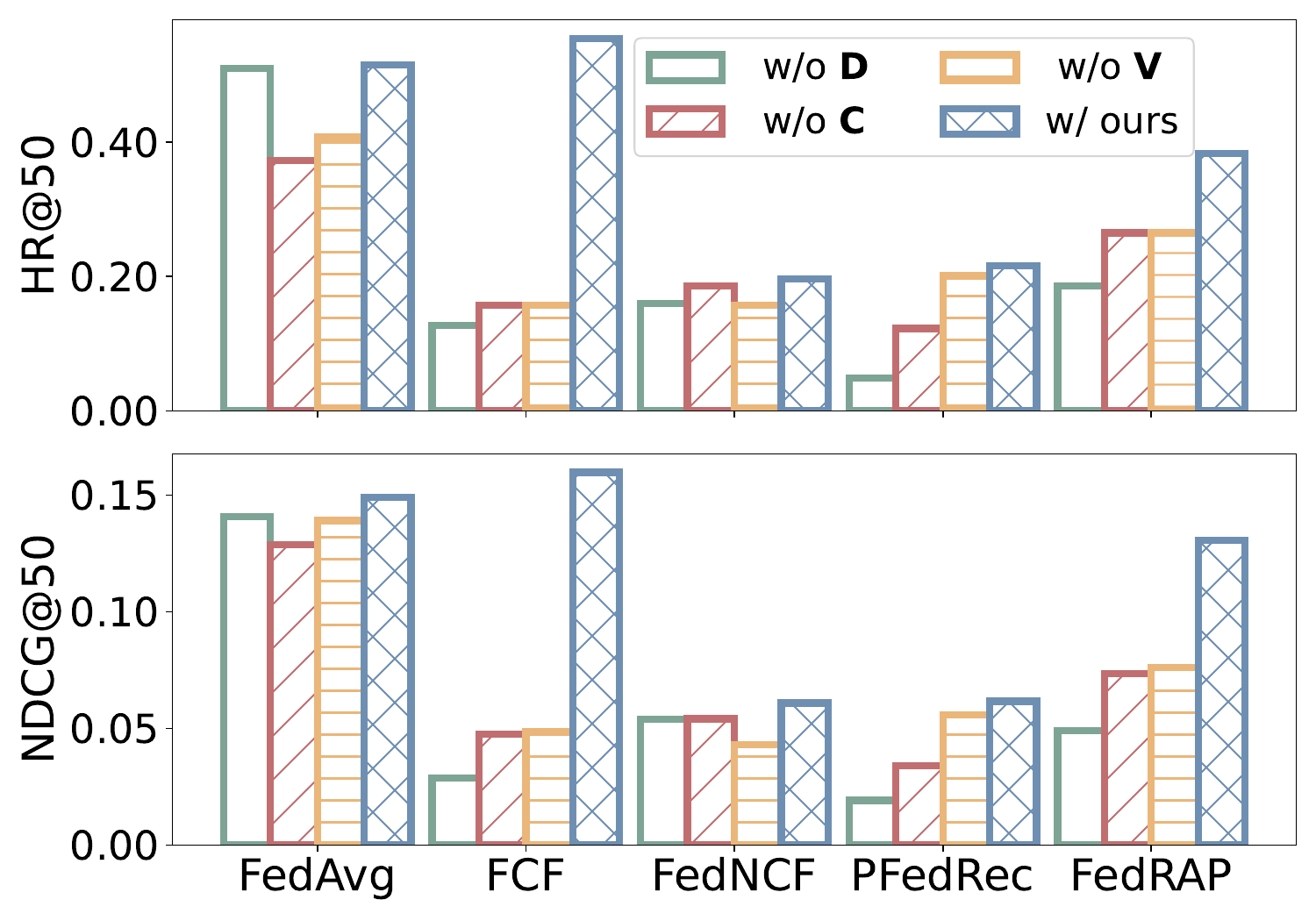}
    \caption{
    Impact on the performance after removing visual (\(\mathbf{V}\)), textual (\(\mathbf{C}\)), or collaborative ID (\(\mathbf{D}\)) features on KU. 
    The varying degrees of the performance degradation across different frameworks demonstrate that there is no universal hierarchy of modality importance, motivating the need of a personalized fusion mechanism for federated VLR tasks.
    }
    \label{fig:modality}
\end{figure}

We analyze the contribution of the visual (\(\mathbf{V}\)), textual (\(\mathbf{C}\)), and collaborative ID (\(\mathbf{D}\)) signals by systematically removing each of them from the model, as shown in Fig.~\ref{fig:modality}. 
Performance consistently degrades when any modality is removed, confirmed that all signals are integral to the accuracy.
The more critical finding is that the relative importance of these modalities is dynamic, depending on the underlying federated architecture. 
For instance, some frameworks are severely impacted by the removal of the collaborative ID signal, while others are more sensitive to the loss of visual or textual content. 
This strong architectural dependency demonstrates that no universal hierarchy of modality importance exists, indicating that the optimal way to combine these diverse signals is highly context-dependent. 
It validates the need for an adaptive fusion mechanism capable of learning this balance.


\begin{table}[!tb]
\centering
{\small
\begin{tabular}{llccc}
\hline
Method                   & Metric & w/ ours   & w/ noise & Degrade              \\ \hline
\multirow{2}{*}{FedAvg}  & HR     & 51.47     & 49.69    & 3.46\% $\downarrow$  \\
                         & NDCG   & 14.91     & 14.16    & 5.03\% $\downarrow$  \\[1pt]
\multirow{2}{*}{FCF}     & HR     & 55.39     & 53.78    & 2.91\% $\downarrow$  \\
                         & NDCG   & 15.97     & 14.02    & 12.21\% $\downarrow$ \\[1pt]
\multirow{2}{*}{FedNCF}  & HR     & 19.61     & 16.80    & 14.33\% $\downarrow$ \\
                         & NDCG   & 6.09      & 4.20     & 31.03\% $\downarrow$ \\[1pt]
\multirow{2}{*}{PFedRec} & HR     & 21.57     & 19.65    & 8.90\% $\downarrow$  \\
                         & NDCG   & 6.16      & 5.86     & 4.87\% $\downarrow$  \\[1pt]
\multirow{2}{*}{FedRAP}  & HR     & 38.24     & 37.25    & 2.59\%  $\downarrow$ \\
                         & NDCG   & 13.05     & 12.36    & 5.29\% $\downarrow$  \\ \hline
\end{tabular}
}
\caption{
Analysis of the privacy–utility trade-off on KU by evaluating the performance impact on the federated baselines enhanced by FedVLR when Gaussian noise is added to the gradients uploaded by clients, comparing results before (\textit{w/ ours}) and after (\textit{w/ noise}) noise addition.
}
\label{table:dp}
\end{table}

\subsubsection{Robustness to Privacy Enhancement.}
We assess FedVLR's compatibility with privacy-enhancing technologies by adding Gaussian noise to the gradients before server aggregation, a common method for achieving differential privacy. 
Table \ref{table:dp} shows that it introduces a predictable and modest trade-off between privacy and utility, with a slight decrease in performance across all models.
Notably, the performance degradation is less severe for simpler or more adaptive methods like FedAvg and FedRAP, suggesting that complex, rigid models may be more sensitive to the perturbations introduced by noise. 
Overall, the results confirm that FedVLR integrates effectively with standard privacy-enhancing mechanisms, demonstrating its practical viability for applications in privacy-conscious environments.

\section{Conclusion}

This paper tackles the challenge of personalized modality fusion in on-device VLR, where uniform fusion logic fails to capture users’ diverse preferences for visual and language content.
We propose FedVLR, a framework with a bi-level fusion mechanism that decouples server-side view generation from lightweight on-device refinement. 
This design empowers each client to learn a fine-grained personalized multimodal fusion module. 
Extensive experiments and theoretical analysis confirm FedVLR substantially improves existing federated baselines, offering a principled pathway toward more personal and privacy-preserving VLR systems.

\bibliography{FedMR}

\include{supp}

\end{document}

%% file: supp.tex
\appendix

\begin{figure*}[t]
    \centering
    \includegraphics[width=.75\linewidth]{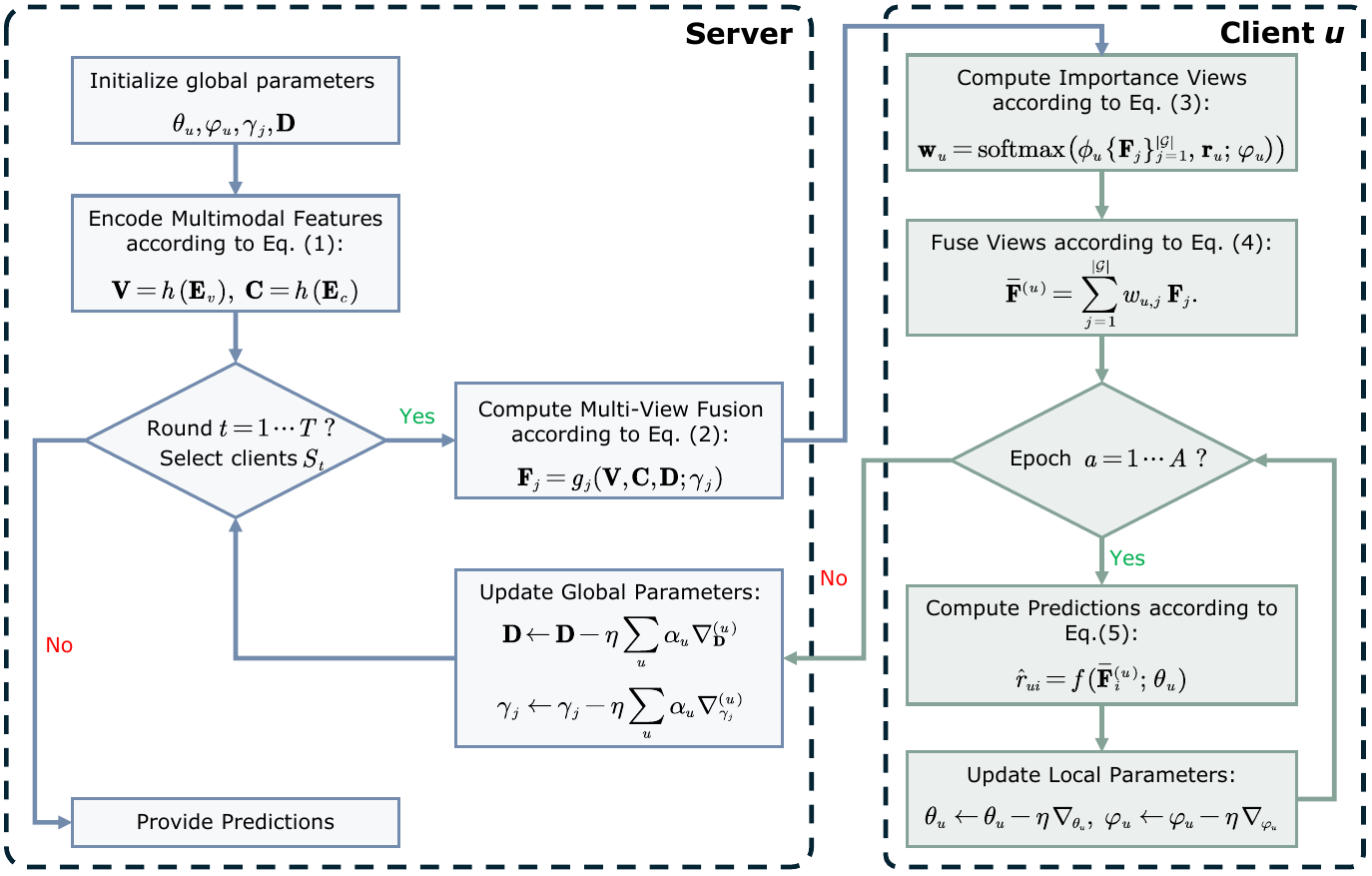}
    \caption{
    Algorithmic workflow of FedVLR, illustrating a clear division of labor: 
    the server offloads heavy computation by generating diverse feature views, while clients perform lightweight on-device personalization. 
    The returned gradients form a collaborative loop that continually refines the global model.
    }
    \label{fig:workflow}
\end{figure*}

\section{Federated Optimization Workflow}

The framework of FedVLRis trained using the iterative workflow shown in Fig.~\ref{fig:workflow}, which is designed around a clear division of labor between the server and clients. 
In each round, the server performs the computationally intensive task of generating diverse feature views \(\{\mathbf{F}_j\}\) from the source modalities \(\mathbf{V}\), \(\mathbf{C}\), and ID embeddings \(\mathbf{D}\). 
By distributing these pre-computed views from the server, it offloads the heavy lifting from the clients. 
This allows each client to focus solely on learning a lightweight personalized refinement model on-device, ensuring that user-specific parameters remain private. 
After local training, clients return gradients for the global components, which the server aggregates to improve its view-generation capabilities. 
This process establishes a collaborative loop where the server learns to create better global views from the collective insights of client-side personalization, and clients in turn benefit from these improved views for more effective training.

\section{Algorithm Optimization}

\subsection{Derivation of the Update for \(\gamma_j^{(u)}\)}

The update process for \(\{\gamma_j^{(u)}\}_{j=1}^{|\mathcal{G}|}\) in Alg. 1 is based on gradient descent. 
The local update at each client \(u\) during the iteration \(t\) follows the following rule:
\begin{equation}
    \gamma_j^{(t)} = \gamma_j^{(t-1)} - \eta \nabla_j^{(u)} \gamma^{(t)},
    \label{eq:update_iter}
\end{equation}
where \(\eta\) is the learning rate, and \(\nabla_j^{(u)} \gamma^{(t)}\) represents the gradient of \(\gamma_j^{(u)}\) at the iteration \(t\) on client \(u\).

To capture how \(\gamma_j^{(u)}\) evolves over time, we can unroll this equation over several iterations. By recursively applying the update rule in Eq. \eqref{eq:update_iter}, we obtain the following:
\begin{equation}
    \gamma_j^{(t)} = \gamma_j^{(0)} - \eta \sum_{i=1}^{t} \nabla_j^{(u)} \gamma^{(i)},
\end{equation}
where \(\gamma_j^{(0)}\) is the initial value of \(\gamma_j^{(u)}\), and the second term is the sum of all gradients computed up to the \(t\)-th iteration. This expression shows that the updated value of \(\gamma_j^{(u)}\) depends on its initial value and the accumulated gradient over \(t\) iterations.

\subsection{Accumulated Gradient}

The accumulated gradient \(\nabla_j^{(u)} \gamma\) at the client \(u\) is the sum of the individual gradients across the iterations, which be expressed as:
\begin{equation}
    \nabla_j^{(u)} \gamma = \frac{1}{\eta} \left( \gamma_j^{(0)} - \gamma_j^{(t)} \right).
    \label{eq:accumluated_gradient}
\end{equation}
Eq. \eqref{eq:accumluated_gradient} represents the total change in \(\gamma_j^{(u)}\) during the local training process. It relates the initial value \(\gamma_j^{(0)}\) to the current value \(\gamma_j^{(t)}\), scaled by the learning rate \(\eta\).

\subsection{Global Aggregation}

In federated learning, the global model is updated by aggregating the local updates from multiple clients. For the parameter \(\gamma_j\), this is achieved by summing the accumulated gradients \(\nabla_j^{(u)} \gamma\) from all selected clients. The global update for \(\gamma_j\) can be written as:
\begin{equation}
    \gamma_j = \gamma_j - \eta \sum_{u \in \mathcal{S}_t} \alpha_u \nabla_j^{(u)} \gamma,
\end{equation}
where \(\mathcal{S}_t\) is the set of clients selected during round \(t\). This step updates the global parameter \(\gamma_j\) by aggregating the contributions from each client’s local updates.
The derived equations demonstrate that the update process is consistent with the principles of gradient-based optimization. This ensures that \(\gamma_j^{(u)}\) is updated in an efficient and mathematically sound manner within the FL framework.

\section{Convergence Analysis}

We provide a theoretical analysis of FedVLR's convergence for the non-convex global objective in Eq.~\eqref{eq:recon}. 
The complete learnable parameter set is denoted as $\mathbf{\Theta} = (\Theta_g, \{\Theta_{l,u}\}_{u \in \mathcal{U}})$, where $\Theta_g = \{\mathbf{D}, \{\gamma_j\}\}$ represents the global parameters and $\Theta_{l,u} = \{\theta_u, \varphi_u\}$ contains the local parameters for client $u$. 
Then the global objective function in Eq.~\eqref{eq:recon} is defined as:
\begin{equation}
\mathcal{J}(\mathbf{\Theta}) = \sum_{u \in \mathcal{U}} \alpha_u \mathcal{J}_u(\mathbf{\Theta}),
\end{equation}
where $\mathcal{J}_u$ is the local objective function on the data of client $u$. Our analysis aims to show that the algorithm finds a stationary point of $\mathcal{J}(\mathbf{\Theta})$. 
The proofs rely on the following standard assumptions used in federated optimization.

\begin{assumption}[L-smoothness]
\label{assump:smooth}
Each local objective $\mathcal{J}_u$ is L-smooth, and the global $\mathcal{J}$ is therefore also L-smooth.
\end{assumption}

\begin{assumption}[Bounded Gradient Variance]
\label{assump:variance}
On each client, the stochastic gradients have a bounded variance:
\begin{equation}
    \mathbb{E}_\xi[\|g_{u}(\mathbf{\Theta};\xi)-\nabla \mathcal{J}_u(\mathbf{\Theta})\|^2]\le\sigma^2.
\end{equation}
The local gradients is also bounded by the expected norm: 
\begin{equation}
    \mathbb{E}_u[\|\nabla \mathcal{J}_u(\mathbf{\Theta})\|^2] \le G^2.
\end{equation}
\end{assumption}

\begin{assumption}[Bounded Heterogeneity]
\label{assump:hetero}
The variance of gradients across users in FedVLR is bounded as: 
\begin{equation}
    \mathbb{E}_{u} [\|\nabla \mathcal{J}_u(\mathbf{\Theta}) - \nabla \mathcal{J}(\mathbf{\Theta})\|^2] \le \zeta^2.
\end{equation}
\end{assumption}

In each communication $t$, clients perform $A$ local steps, leading to a drift between the locally updated parameters and the global model state. 
This local drift is a key challenge in analyzing federated algorithms. We first bound this drift:

\begin{lemma}[Local Parameter Drift]
\label{lemma:drift}
Let $\mathbf{\Theta}_{l,u}^{t,a}$ be the local parameters on user $u$ after $a$ local steps in round $t$. 
Under our assumptions, the expected drift after $A$ steps is bounded:
\begin{equation}
\mathbb{E} \left[ \|\mathbf{\Theta}_{l,u}^{t,A} - \mathbf{\Theta}_{l,u}^{t,0} \|^2 \right] \le A^2 \eta^2 G^2.
\end{equation}
\end{lemma}

\begin{proof}[Proof of Lemma~\ref{lemma:drift} (Local Update Bound)]

Using the local update rule:
\begin{equation}
\Theta_{l,u}^{t,k} = \Theta_{l,u}^{t,k-1} - \eta\, g_{l,u}(\Theta_{l,u}^{t,k-1};\xi),
\end{equation}
and Assumption 2, we have:
\begin{equation}
\begin{split}
&\mathbb{E}\Bigl\|\Theta_{l,u}^{t,k} - \Theta_{l,u}^t\Bigr\|^2 \\
= &\mathbb{E}\left\|\sum_{i=1}^k \eta\, g_{l,u}(\Theta_{l,u}^{t,i-1};\xi)\right\|^2 \\[4pt]
\leq & k\,\eta^2 \sum_{i=1}^k \mathbb{E}\left\|g_{l,u}(\Theta_{l,u}^{t,i-1};\xi)\right\|^2 
\quad \text{(by Jensen's inequality)}\\[4pt]
\leq & k^2\,\eta^2 (G^2 + \sigma_l^2) \quad \text{(by Assumption 2).}
\end{split}
\end{equation}

Jensen's inequality is used because the square of a sum under an expectation can be bounded by \(k\) times the sum of squares. 

For \(\eta = \tfrac{1}{L\sqrt{A}}\) and \(k = A\), it follows that:
\begin{equation}
\begin{split}
\mathbb{E}\Bigl\|\Theta_{l,u}^{t,A} - \Theta_{l,u}^t\Bigr\|^2 
&\leq A^2\,\eta^2 (G^2 + \sigma_l^2)
= \frac{A(G^2 + \sigma_l^2)}{L^2}.
\end{split}
\end{equation}

Defining \(C_1 = A\), we can equivalently express the bound as the followings:
\begin{equation}
\begin{split}
\mathbb{E}\Bigl\|\Theta_{l,u}^{t,A} - \Theta_{l,u}^t\Bigr\|^2 
\leq C_1\,A\,\eta^2 (G^2 + \sigma_l^2).
\end{split}
\end{equation}
\end{proof}

The local drift leads to a bias in the aggregated gradients uploaded to the server. 
We then bounds the discrepancy between the averaged gradient and the true global gradient:

\begin{lemma}[Gradient Bias Bound]
\label{lemma:bias}
Let $\bar{g}_g^t$ be the averaged gradient for the global parameters $\Theta_g$ after $A$ local updates. 
The expected squared norm of the gradient bias is bounded:
\begin{equation}
\mathbb{E}[\|\mathbb{E}[\bar{g}_g^t | \mathbf{\Theta}^t] - \nabla_{\Theta_g} \mathcal{J}(\mathbf{\Theta}^t)\|^2] \le L^2 A^2 \eta^2 G^2.
\end{equation}
\end{lemma}

\begin{proof}[Proof of Lemma~\ref{lemma:bias} (Gradient Bias Bound)]

The gradient bias is given by
\begin{equation}
\begin{split}
\|B_t\|
&= \Bigl\|\mathbb{E}_u\Bigl[\nabla_{\Theta_g}F_u\bigl(\Theta_g^t,\Theta_{l,u}^{t,A}\bigr)\Bigr] - \nabla_{\Theta_g}F(\Theta_g^t)\Bigr\| \\[4pt]
&= \Bigl\|\mathbb{E}_u\Bigl[\nabla_{\Theta_g}F_u\bigl(\Theta_g^t,\Theta_{l,u}^{t,A}\bigr) - \nabla_{\Theta_g}F_u\bigl(\Theta_g^t,\Theta_{l,u}^t\bigr)\Bigr]\Bigr\| \\[4pt]
&\leq \mathbb{E}_u\Bigl\|\nabla_{\Theta_g}F_u\bigl(\Theta_g^t,\Theta_{l,u}^{t,A}\bigr) - \nabla_{\Theta_g}F_u\bigl(\Theta_g^t,\Theta_{l,u}^t\bigr)\Bigr\| \\[4pt]
&\leq L\, \mathbb{E}_u\Bigl\|\Theta_{l,u}^{t,A} - \Theta_{l,u}^t\Bigr\|,
\end{split}
\end{equation}
where the last inequality follows from Assumption 1. Squaring both sides and applying the result from Lemma 1 yields:
\begin{equation}
\begin{split}
\|B_t\|^2 
\leq L^2\, \mathbb{E}_u\Bigl\|\Theta_{l,u}^{t,A} - \Theta_{l,u}^t\Bigr\|^2 
\leq L^2\, C_1\, A\,\eta^2 (G^2 + \sigma_l^2).
\end{split}
\end{equation}
\end{proof}

\begin{lemma}[Aggregated Gradient Second Moment Bound]
\label{lemma:second_moment}
Under Assumptions 2, 3, and Lemma 2, assuming uniform client sampling on \(n_s\) clients, we can derive the aggregated gradient second moment bound as followings:
\begin{equation}
\mathbb{E}[\|\bar{g}_g^t\|^2] \le 2\mathbb{E}[\|\nabla F(\Theta_g^t)\|^2] + 2\mathbb{E}[\|B_t\|^2] + \frac{2(\sigma^2+G^2)}{n_s}
\end{equation}
\end{lemma}

\begin{proof}[Proof of Lemma \ref{lemma:second_moment} (Aggregated Gradient Second Moment)]

When Assumptions 2 and 3 hold, we can use the variance decomposition to derive the followings:
\begin{equation}
\small
\begin{split}
\mathbb{E}\Bigl\|\bar{g}_g^t\Bigr\|^2 
&= \Bigl\|\mathbb{E}\bar{g}_g^t\Bigr\|^2 + \mathbb{E}\Bigl\|\bar{g}_g^t - \mathbb{E}\bar{g}_g^t\Bigr\|^2 \\[3pt]
&= \Bigl\|\nabla F(\Theta_g^t) + B_t\Bigr\|^2 + \operatorname{Var}\left(\bar{g}_g^t\right) \\[3pt]
&\leq 2\Bigl\|\nabla F(\Theta_g^t)\Bigr\|^2 + 2\|B_t\|^2 + \frac{1}{n_s}\,\mathbb{E}_u\Bigl\|\nabla_{\Theta_g}F_u\bigl(\Theta_g^t,\Theta_{l,u}^{t,A}\bigr)\Bigr\|^2 \\[3pt]
&\leq 2\Bigl\|\nabla F(\Theta_g^t)\Bigr\|^2 + 2\|B_t\|^2 + \frac{G^2 + \sigma^2}{n_s}.
\end{split}
\end{equation}
The last step is based on Assumptions 2 and 3. 
\end{proof}

\begin{proof}[Proof of Theorem 1 (Convergence Rate)]

Starting from the \(L\)-smoothness of \(F\), which is in Assumpution 1, we have:
\begin{equation}
\begin{split}
F(\Theta_g^{t+1}) 
&\leq F(\Theta_g^t) + \bigl\langle \nabla F(\Theta_g^t), \Theta_g^{t+1}-\Theta_g^t \bigr\rangle 
+ \frac{L}{2}\Bigl\|\Theta_g^{t+1}-\Theta_g^t\Bigr\|^2 \\[3pt]
&= F(\Theta_g^t) - \eta\,\bigl\langle \nabla F(\Theta_g^t), \bar{g}_g^t \bigr\rangle 
+ \frac{L\eta^2}{2}\Bigl\|\bar{g}_g^t\Bigr\|^2.
\end{split}
\end{equation}

Taking expectations and using Lemmas 2 and 3, we obtain:
\begin{equation}
\small
\begin{split}
\mathbb{E}F(\Theta_g^{t+1})
&\leq \mathbb{E}F(\Theta_g^t) 
- \eta\,\mathbb{E}\Bigl\langle \nabla F(\Theta_g^t), \nabla F(\Theta_g^t) + B_t \Bigr\rangle 
+ \frac{L\eta^2}{2}\,\mathbb{E}\Bigl\|\bar{g}_g^t\Bigr\|^2 \\[3pt]
&\leq \mathbb{E}F(\Theta_g^t) 
- \frac{\eta}{2}\,\mathbb{E}\Bigl\|\nabla F(\Theta_g^t)\Bigr\|^2 
+ \frac{\eta}{2}\,\mathbb{E}\|B_t\|^2 \\[3pt]
&\quad + L\eta^2\,\mathbb{E}\Bigl\|\nabla F(\Theta_g^t)\Bigr\|^2 
+ L\eta^2\,\mathbb{E}\|B_t\|^2 
+ \frac{L\eta^2(G^2+\sigma^2)}{2n_s}.
\end{split}
\end{equation}

Here we see the interplay of the gradient norm \(\|\nabla F(\Theta_g^t)\|\) and the bias term \(B_t\). We control these via the learning rate \(\eta\). 
Choosing \(\eta \leq \frac{1}{4L}\) ensures certain terms are upper-bounded by \(\frac{1}{2}\|\nabla F(\Theta_g^t)\|^2\).

For \(\eta \leq \frac{1}{4L}\), collecting like terms yields:
\begin{equation}
\small
\begin{split}
\frac{\eta}{4}\,\mathbb{E}\Bigl\|\nabla F(\Theta_g^t)\Bigr\|^2 
\leq & \mathbb{E}\Bigl[F(\Theta_g^t) - F(\Theta_g^{t+1})\Bigr] \\
+ & \frac{5L^3 C_1 A\,\eta^3(G^2+\sigma_l^2)}{2} \\
+ &\frac{L\eta^2(G^2+\sigma^2)}{2n_s}.
\end{split}
\end{equation}

Summing over \(t=0\) to \(T-1\) and setting \(\eta = \frac{1}{L\sqrt{T}}\), we obtain
\begin{equation}
\small
\begin{split}
\frac{1}{T}\sum_{t=0}^{T-1}\mathbb{E}\Bigl\|\nabla F(\Theta_g^t)\Bigr\|^2 
&\leq \frac{4L\,(F_0-F^*)}{\sqrt{T}} 
+ \frac{10C_1 A G^2}{T} 
+ \frac{2(G^2+\sigma^2)}{\sqrt{T}\,n_s}.
\end{split}
\end{equation}
The additional term \(\mathcal{O}\bigl(G^2/T\bigr)\) is dominated by
\(\mathcal{O}\bigl(G^2/\sqrt{T}\bigr)\), which concludes the proof.
\end{proof}

\section{More Experiments}

\subsection{More Experiment Details}

All baseline methods are implemented in PyTorch \cite{paszke2019pytorch}, and their official implementations are publicly available. Experiments are conducted on a machine with a 3.60 GHz Intel Xeon Gold 6244 CPU, three Quadro RTX 6000 GPUs, and 395 GB of RAM.

\subsection{Convergence Analysis}
\label{sec:convergence_analysis}

\begin{figure}[!tbp]
    \centering
    \includegraphics[width=.95\linewidth]{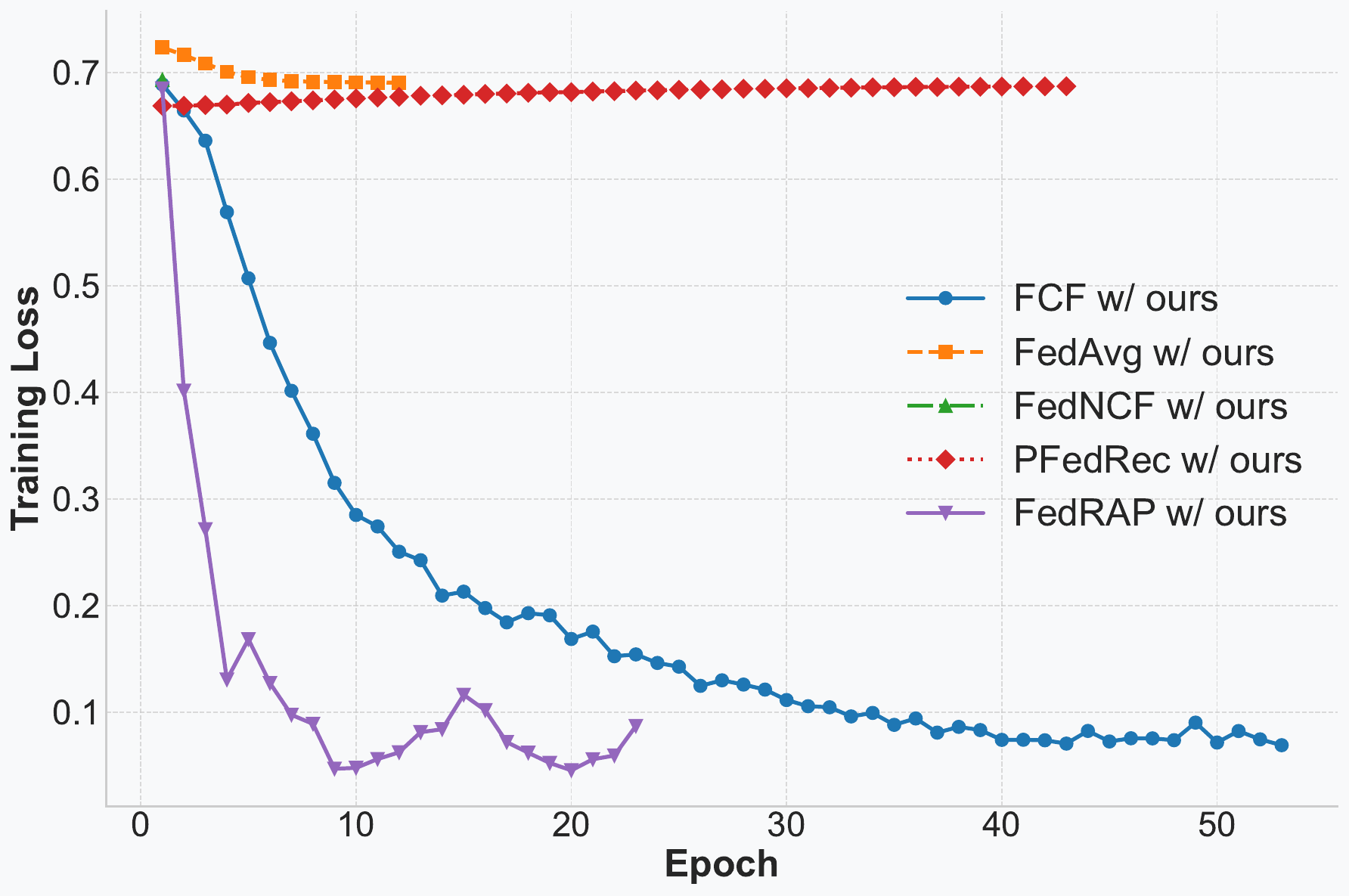}
    \caption{Training loss progression over epochs for baseline methods enhanced with FedVLR on the KU dataset, illustrating convergence behavior.}
\label{fig:training_loss}
\end{figure}

To assess the training stability and convergence properties of the baseline methods when enhanced with FedVLR, we plot their training loss progression over 50 epochs on the KU dataset in Figure~\ref{fig:training_loss}. 
The loss curves demonstrate that all FedVLR-enhanced models exhibit stable convergence behavior. 
The training loss decreases consistently across epochs for all methods, with a sharp reduction typically observed in the initial epochs followed by more gradual convergence towards a stable value. 
This indicates that the FedVLR framework integrates well with diverse on-device backbone architectures, allowing for effective optimization and stable learning dynamics. 
While the final converged loss values vary slightly across methods, all models successfully minimize the training objective, confirming the robustness of the training process when incorporating FedVLR's multimodal fusion capabilities.

\begin{figure*}[!tb]
    \centering
    \includegraphics[width=1\linewidth]{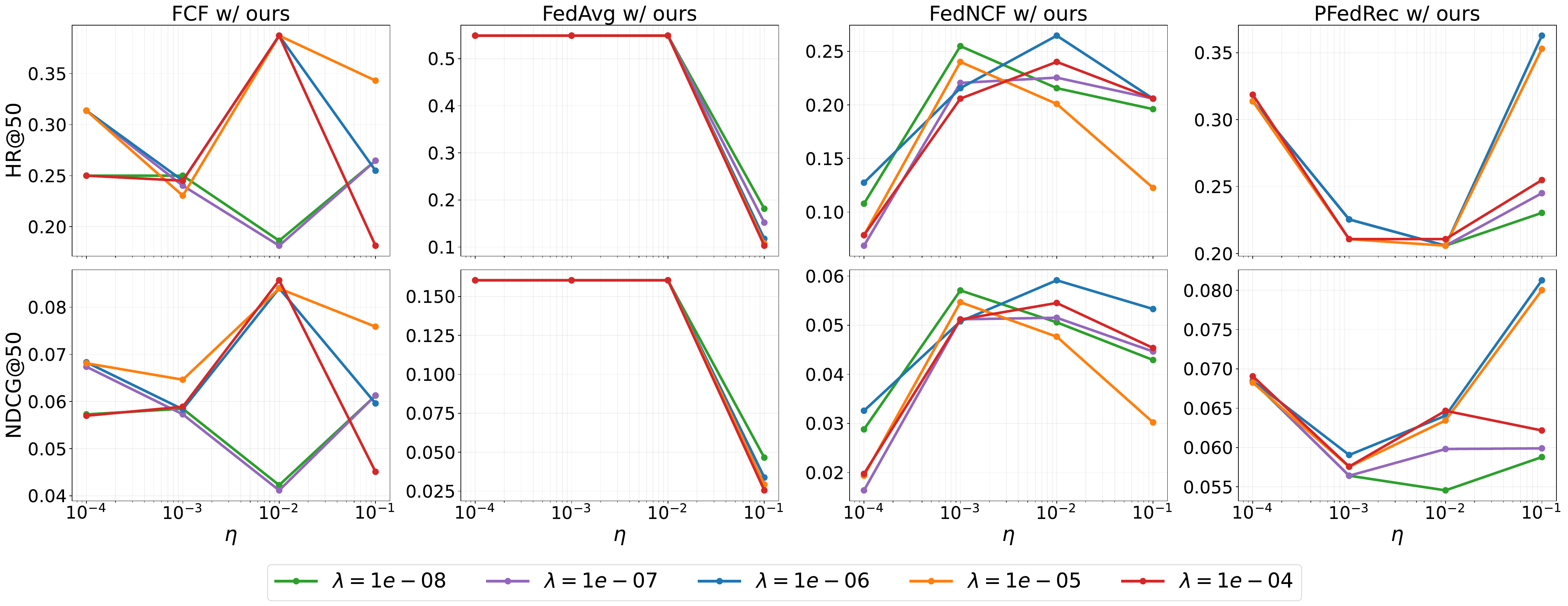}
    \caption{Sensitivity analysis of four FedVLR-enhanced baselines (FCF, FedAvg, FedNCF, PFedRec) on the KU dataset. Performance is plotted against varying learning rates \(\eta\) and L2 regularization  \(\lambda\) strengths.}
    \label{fig:hyper}
\end{figure*}

\begin{figure*}[!tb]
    \centering
    \includegraphics[width=1\linewidth]{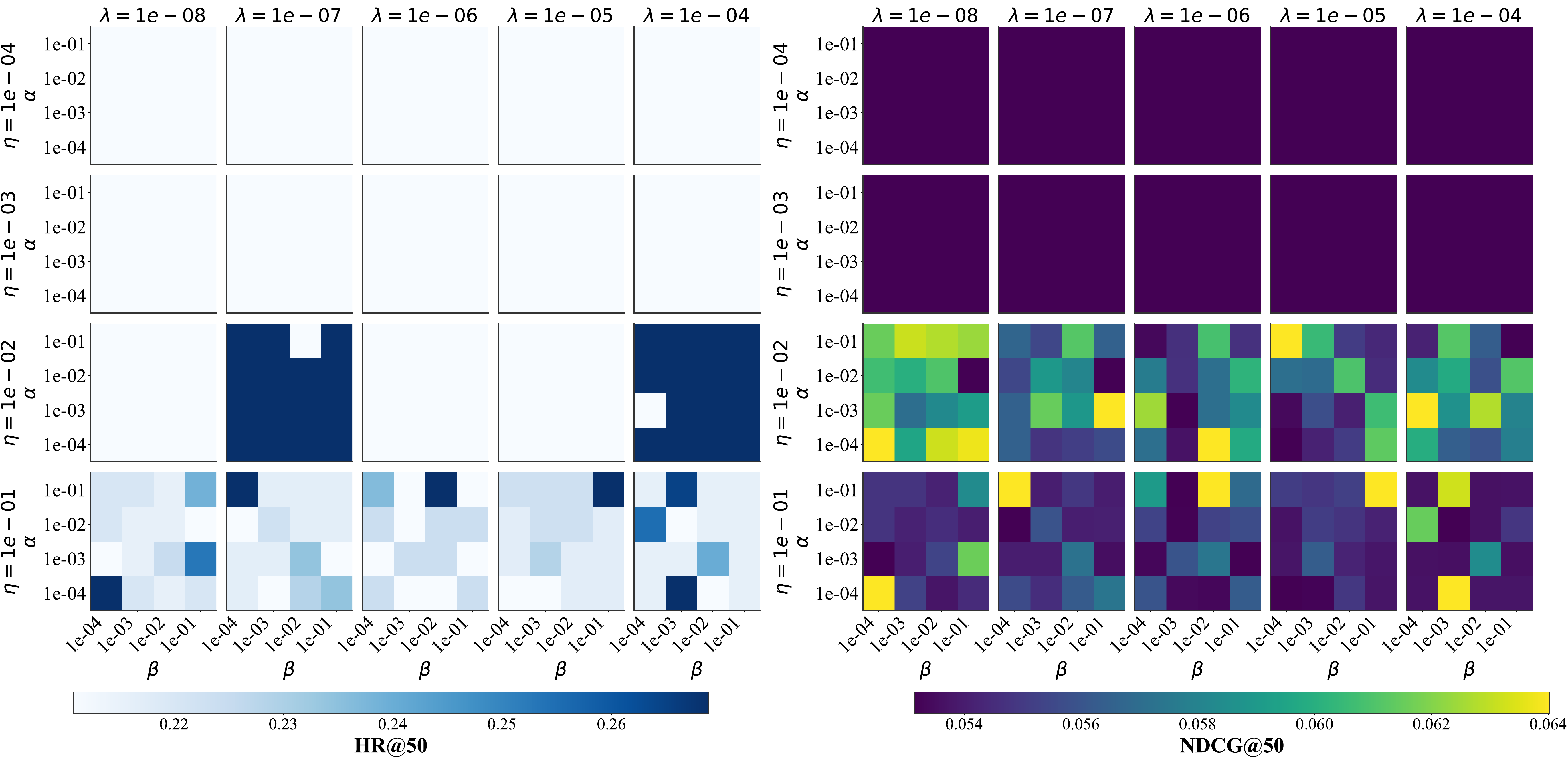}
    \caption{Hyperparameter tuning for FedVLR-enhanced FedRAP on the KU dataset. Heatmaps visualize HR@50 and NDCG@50 across personalization parameters \(\alpha\) and \(\beta\) for different learning rates \(\eta\) and L2 regularization  \(\lambda\). 
    Performance intensity is indicated by color.
    }
    \label{fig:hyper_fedrap}
\end{figure*}

\subsection{Hyperparameter Analysis}
\label{sec:hyperparameter_analysis}

We analyze the sensitivity of FedVLR-enhanced models to key hyperparameters, specifically the learning rate \(\eta\) and L2 regularization \(\lambda\) strength, using the KU dataset. 
Fig.~\ref{fig:hyper} illustrates this for FedVLR applied to FCF, FedAvg, FedNCF, and PFedRec. 
Across these models, the learning rate exhibits a significant impact, with performance generally peaking at intermediate values and decreasing notably at lower or higher rates. 
This highlights the necessity of selecting an appropriate learning rate for effective convergence. 
The influence of L2 regularization appears moderate in comparison, though still relevant. 
No single \(\lambda\) value is universally optimal, with the best performance achieved at different strengths depending on the backbone and learning rate, necessitating model-specific tuning.

The tuning process for the FedVLR-enhanced FedRAP model, which includes its specific personalization parameters \(\alpha\) and \(\beta\), is detailed in Fig.~\ref{fig:hyper_fedrap}. The heatmaps clearly indicate high sensitivity to the choices of \(\alpha\) and \(\beta\), with distinct optimal regions emerging. Furthermore, these optimal regions shift based on the learning rate, signifying a strong interaction between \(\eta\), \(\alpha\), and \(\beta\). This underscores the need for careful co-tuning, typically via grid search, to identify the best configuration for FedVLR-enhanced FedRAP. Overall, while the models demonstrate reasonable robustness, achieving optimal results requires careful hyperparameter tuning, particularly for the learning rate and any model-specific parameters. The best hyperparameters identified through validation during this analysis informed the settings used for the main experimental results.

\begin{figure}[tb]
    \centering
    \includegraphics[width=0.95\linewidth]{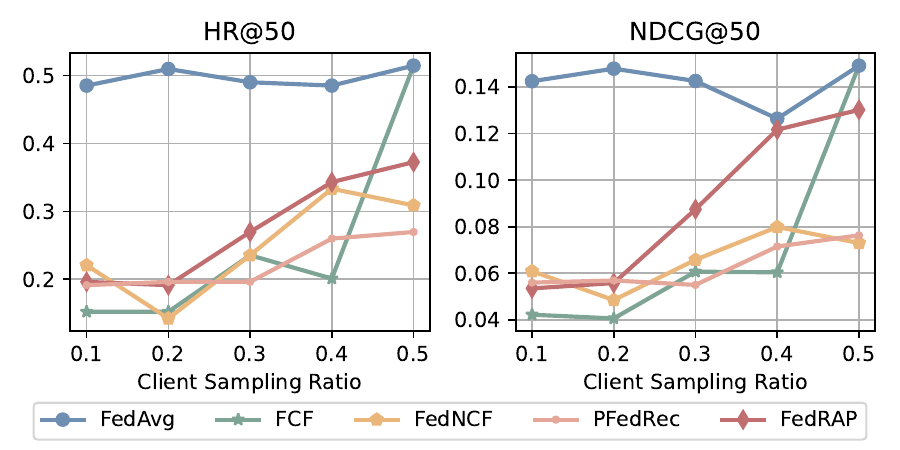}
    \caption{Performance comparsions of different federated methods with FedVLR under varying client sampling ratios on HR@50 and NDCG@50.}
    \label{fig:ab_client}
\end{figure}

\subsection{Ablation Study on Client Sampling}

To investigate the impact of introducing FedVLR for multimodal data fusion on different federated methods under varying client sampling ratios, we conduct experiments on the five on-device methods using the KU dataset. 
Fig. \ref{fig:ab_client} presents the results for HR@50 and NDCG@50 as the client sampling ratio changes. 
Overall, as the client sampling ratio increases, most methods show improvements in HR@50 and NDCG@50, indicating that more client participation helps enhance the personalization of recommendations.
However, the performance improvement varies across different methods. 
For PFedRec, the increase in HR@50 and NDCG@50 after introducing FedVLR is relatively small, possibly due to its lower sensitivity to the number of participating clients. 
This suggests that PFedRec may rely more on the quality of individual client data rather than the number of clients. 
As a result, the advantages of FedVLR's multimodal fusion are not fully realized when the client count is low. 
Additionally, when the client sampling ratio is low, the model's performance becomes unstable, indicating that insufficient data affects the stability of feature learning. 
This is especially true in multimodal scenarios, where a lack of sufficient modality information weakens the fusion effect, reducing recommendation accuracy.

\begin{figure*}[t]
    \centering
    \includegraphics[width=.95\linewidth]{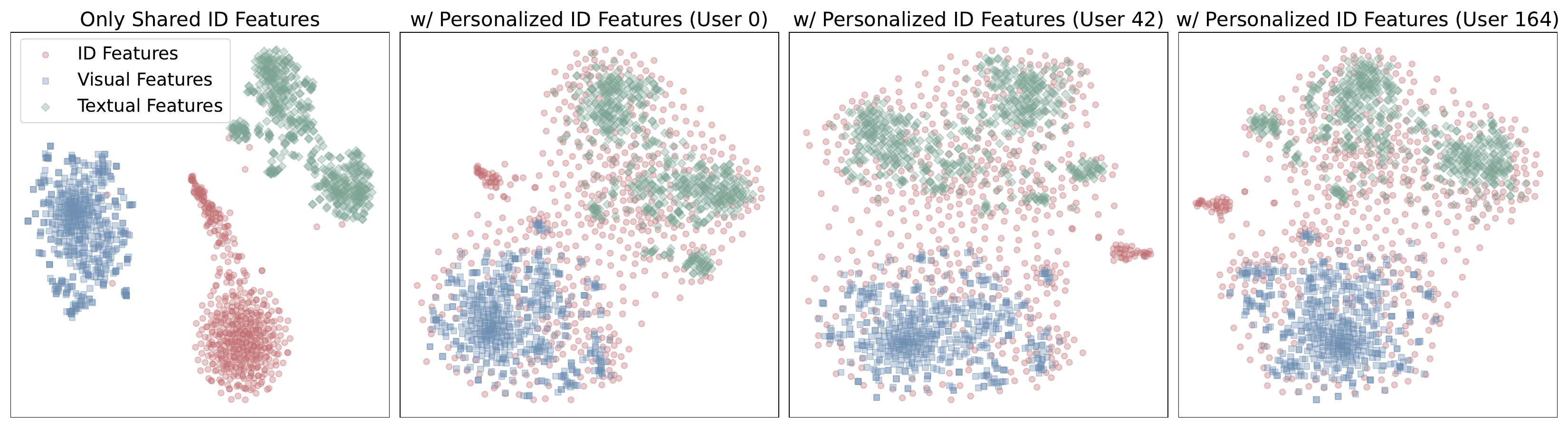}
    \caption{Visualization of multimodal features and ID features mapped into a shared representation space.}
    \label{fig:ab_modalities}
\end{figure*}

\subsection{Ablation Study on Modality Distributions}

To analyze the interaction between multimodal features and ID features, we use FedRAP as the backbone for FedVLR. Fig. \ref{fig:ab_modalities} visualizes the distribution of these features on the KU dataset. 
The first subfigure shows that globally shared ID features and multimodal features occupy distinct subregions in the shared representation space, a phenomenon known as the "Modality Gap" \cite{liang2022mind,zhang2023diagnosing}. 
This separation enhances the robustness of multimodal data processing by reducing noise interference. 
The subsequent subfigures illustrate how personalized ID features influence this distribution. Incorporating personalized ID embeddings (e.g., for Users \(0\) and \(42\)) causes the ID features to expand toward the regions of visual and textual features, effectively shrinking the modality gap. This adjustment facilitates multimodal fusion and improves the accuracy of personalized recommendations. 
By comparing users, it is evident that the model dynamically adapts feature distributions based on user-specific behaviors and preferences, better addressing individual needs. This demonstrates the critical role of personalized ID features in federated recommendation systems, enabling effective personalization while preserving user privacy.

\begin{table*}[t]
\centering
{\setlength{\tabcolsep}{.85mm}\small
\begin{tabular}{clccccccc}
\hline
Method                   & Metric  & w/ ours  & w/o \(\mathbf{V}\)                             & w/ rnd \(\mathbf{V}\)                          & w/o \(\mathbf{C}\)                             & w/ rnd \(\mathbf{C}\)                          & w/o \(\mathbf{D}\)                            & w/ rnd \(\mathbf{D}\)                          \\ \hline
\multirow{2}{*}{FedAvg}  & HR   & 51.47 & 40.69 (20.94\% \(\downarrow\)) & 39.22 (23.80\% \(\downarrow\)) & 37.25 (27.63\% \(\downarrow\)) & 36.76 (28.58\% \(\downarrow\)) & 50.94 (1.03\% \(\downarrow\))  & 45.20 (12.18\% \(\downarrow\)) \\
                         & NDCG & 14.91 & 13.91 (6.71\% \(\downarrow\))  & 13.03 (12.61\% \(\downarrow\)) & 12.87 (13.68\% \(\downarrow\)) & 12.40 (16.83\% \(\downarrow\)) & 14.07 (5.66\% \(\downarrow\))  & 13.75 (7.78\% \(\downarrow\))  \\[3pt]
\multirow{2}{*}{FCF}     & HR   & 55.39 & 28.43 (48.67\% \(\downarrow\)) & 28.92 (47.79\% \(\downarrow\)) & 42.16 (23.89\% \(\downarrow\)) & 24.51 (55.75\% \(\downarrow\)) & 23.53 (57.52\% \(\downarrow\)) & 22.06 (60.17\% \(\downarrow\)) \\
                         & NDCG & 15.97 & 7.48 (53.16\% \(\downarrow\))  & 7.83 (50.97\% \(\downarrow\))  & 9.65 (39.57\% \(\downarrow\))  & 6.29 (60.61\% \(\downarrow\))  & 5.78 (63.81\% \(\downarrow\))  & 5.38 (66.31\% \(\downarrow\))  \\[3pt]
\multirow{2}{*}{FedNCF}  & HR   & 19.61 & 15.69 (19.99\% \(\downarrow\)) & 16.67 (14.99\% \(\downarrow\)) & 18.63 (5.00\% \(\downarrow\))  & 16.18 (17.49\% \(\downarrow\)) & 16.02 (18.31\% \(\downarrow\)) & 15.69 (19.99\% \(\downarrow\)) \\
                         & NDCG & 6.09  & 4.31 (29.23\% \(\downarrow\))  & 4.89 (19.70\% \(\downarrow\))  & 5.41 (11.17\% \(\downarrow\))  & 4.61 (24.30\% \(\downarrow\))  & 5.38 (11.66\% \(\downarrow\))  & 4.58 (24.79\% \(\downarrow\))  \\[3pt]
\multirow{2}{*}{PFedRec} & HR   & 21.57 & 19.90 (7.74\% \(\downarrow\))  & 16.47 (23.64\% \(\downarrow\)) & 19.41 (10.01\% \(\downarrow\)) & 16.96 (21.37\% \(\downarrow\)) & 17.94 (16.83\% \(\downarrow\)) & 16.47 (23.64\% \(\downarrow\)) \\
                         & NDCG & 6.16  & 4.69 (23.86\% \(\downarrow\))  & 4.68 (24.03\% \(\downarrow\))  & 5.37 (12.82\% \(\downarrow\))  & 4.45 (27.76\% \(\downarrow\))  & 4.73 (23.21\% \(\downarrow\))  & 4.84 (21.43\% \(\downarrow\))  \\[3pt]
\multirow{2}{*}{FedRAP}  & HR   & 38.24 & 26.47 (30.78\% \(\downarrow\)) & 22.06 (42.31\% \(\downarrow\)) & 32.35 (15.40\% \(\downarrow\)) & 23.04 (39.75\% \(\downarrow\)) & 18.63 (51.29\% \(\downarrow\)) & 13.24 (65.38\% \(\downarrow\)) \\
                         & NDCG & 13.05 & 7.62 (41.61\% \(\downarrow\))  & 6.34 (51.42\% \(\downarrow\))  & 7.14 (45.29\% \(\downarrow\))  & 5.64 (56.78\% \(\downarrow\))  & 4.90 (62.43\% \(\downarrow\))  & 3.38 (74.10\% \(\downarrow\))  \\ \hline
\end{tabular}
}
\caption{Performance comparison on the KU dataset, reporting HR@50 (HR) and NDCG@50 (NDCG) in percentage for several federated baslines enhanced with our FedVLR. 
Percentages indicate the performance drop relative to the full-feature baseline.}

\label{table:modality_complete}
\end{table*}

\subsection{Ablation Study on Modality Contribution}
\label{sec:ablation_modality}

We conduct an ablation study to analyze the contribution of each modality to the final performance. 
As shown in Table~\ref{table:modality_complete}, all three information sources are essential for effective prediction. 
Removing any single modality, whether visual (\(\mathbf{V}\)), textual (\(\mathbf{C}\)), or collaborative ID (\(\mathbf{D}\)), consistently degrades performance across all federated architectures. 
Replacing a modality with random noise generally causes an even larger performance drop than removing it entirely, suggesting the model actively learns meaningful patterns from the semantic content of each modality. 
Injecting noise is more detrimental than a missing signal because the model must expend capacity to learn robustly to ignore the uninformative features.

The most critical insight from this analysis is that there is no universal hierarchy of modality importance. The degree of performance degradation from removing a specific modality is highly dependent on the underlying architecture of the federated baseline. For instance, some architectures are severely impacted by the removal of the collaborative ID signal, while others are more sensitive to the loss of visual or textual content. This strong architectural dependency shows that the optimal way to combine multimodal signals is not fixed. It provides a powerful motivation for our proposed method. If different model architectures require different fusion priorities, it follows that individual users with diverse tastes will also benefit from a personalized fusion mechanism that can adaptively balance these information sources.

\subsection{Ablation Study on Fusion Strategy}

Table \ref{tb:ab_experts_hr} and Table \ref{tb:ab_experts_ndcg} shows the HR@50 and NDCG@50 results for five FedRec methods with different fusion strategies on the KU dataset. 
The results highlight the critical role of fusion strategies in performance. 
Among them, \textbf{Ours} strategy, which dynamically combines fusion methods based on user interaction history, achieves the best performance.
In contrast, the \textbf{Sum} strategy performs the worst across all metrics.
The \textbf{Sum} strategy, which simply adds features, fails to account for complex interactions among modalities, often leading to information loss and poor handling of long-tail distributions. 
In contrast, the \textbf{MLP} strategy better captures nonlinear interactions but is limited by its network depth and parameter count. 
The \textbf{Gate} strategy dynamically learns feature weights, optimizing the importance of different modalities. 
However, \textbf{Ours} strategy combines all fusion methods, dynamically adjusting their contributions, enabling it to generate fine-grained personalized item representations and better capture complex relationships among features.
This study underscores the importance of personalized item representations in FedMMRec. 
By leveraging the BLFM, FedVLR allows each client to dynamically adjust fusion weights based on user interaction history, enabling fine-grained personalization and enhancing recommendation accuracy while maintaining privacy under the federated settings. \looseness=-1

\begin{table}[t]
\small
\begin{tabular}{cccccc}
\hline
Expert & FedAvg & FCF    & FedNCF & PFedRec & FedRAP \\ \hline
Sum    & 0.1814 & 0.2255 & 0.2353 & 0.1912  & 0.2353 \\
MLP    & 0.3627 & 0.5098 & 0.2402 & 0.0637  & 0.3333 \\
Gate   & 0.5139 & 0.2794 & 0.1667 & 0.1275  & 0.2304 \\
Ours    & 0.5147 & 0.5539 & 0.1961 & 0.2157  & 0.3824 \\ \hline
\end{tabular}
\caption{Performance comparison on HR@50 of different fusion strategies integrated with FedVLR on KU.}
\label{tb:ab_experts_hr}
\end{table}

\begin{table}[t]
\small
\begin{tabular}{cccccc}
\hline
Expert & FedAvg & FCF    & FedNCF & PFedRec & FedRAP \\ \hline
Sum    & 0.0570 & 0.0660 & 0.0529 & 0.0390  & 0.0639 \\
MLP    & 0.1215 & 0.1250 & 0.0538 & 0.0159  & 0.0856 \\
Gate   & 0.1397 & 0.0741 & 0.0443 & 0.0283  & 0.0604 \\
Ours    & 0.1491 & 0.1597 & 0.0609 & 0.0616  & 0.1305 \\ \hline
\end{tabular}
\caption{Performance comparison on NDCG@50 of different fusion strategies integrated with FedVLR on KU.}
\label{tb:ab_experts_ndcg}
\end{table}